# The Influence of Synoptic Wind on Coastal Circulation Dynamics

Mohammad Allouche[a], Elie Bou-Zeid[a,*], Juho Iipponen[b]

[a] *Department of Civil and Environmental Engineering, Princeton University, New Jersey*

[b] *Program in Atmospheric and Oceanic Sciences, Princeton University, New Jersey*

*Corresponding author: Elie Bou-Zeid ebouzeid@princeton.edu

## ABSTRACT

Particularly challenging classes of heterogeneous surfaces are ones where strong secondary circulations are generated, potentially dominating the flow dynamics. In this study, we focus on land-sea breeze circulations (LSBs) resulting from surface thermal contrasts in the presence of increasing synoptic pressure forcing. The relative importance and orientation of the thermal and synoptic forcings are measured through two dimensionless parameters: a bulk Richardson number $Ri = W_*^2/M_g^2$ ($M_g$ is the geostrophic wind magnitude and $W_*$ a convective buoyant velocity scale), and the angle $\alpha$ between the shore and geostrophic wind. Large eddy simulations reveal the emergence of various regimes where the dynamics are shown to be asymmetric with respect to $\alpha$. Along-shore cases result in deep LSBs similar to the quiescent scenario (zero synoptic background), irrespective of the strength of $M_g$. Across-shore simulations exhibit a circulation cell that decreases in height with increasing synoptic forcing. However, at the highest synoptic winds simulated, the circulation cell is advected away with sea-to-land winds, while a shallow circulation persists for land-to-sea cases. Scaling analysis that relates the internal parameters $Q_{shore}$ (net shore volumetric flux) and $q_{shore}$ (net shore advected heat flux) to the external input parameters $M_g$ and $W_*$ results in a succinct model of the shore fluxes that also helps explain the physical implications of the identified LSBs. Finally, the vertical profiles of the shore-normal velocity and shore-advected heat flux are used, with the aid of $k$-means clustering, to independently classify the LSBs in the ($Ri^{-1}$, $\alpha$) regime space diagram (canonical LSB, sea-driven LSB, land-driven LSB, and advected LSB), corroborating our visual categorization.

# 1   Introduction

Land surface heterogeneity remains a pivotal open challenge in boundary layer meteorology as it hinders the accurate forecasting of micro to synoptic scale atmospheric dynamics. Specifically, heterogeneity plays a key role in marine boundary layers (LaCasse *et al.*, 2008; Sullivan *et al.*, 2020, 2021) and coastal zones where predicting the climate, the weather, and air pollution dispersion is most critical (Rotunno *et al.*, 1996; Miller *et al.*, 2003). Coastal circulation dynamics are of particular interest since they have a direct impact on coastal zones that house over 50% of the world's population and 70% of its megacities (D. Hinrichsen, 1997; Crosman and Horel, 2010a), and these zones will be increasingly vulnerable to natural hazards that will be exacerbated by climate change (hurricanes, sea level rise, heat waves).

As reviewed in (Bou-Zeid *et al.*, 2020), the LSBs that arise in these zones belong to a more general class of heterogeneity problems (the semi-infinite interfaces) that has been studied extensively in the literature theoretically (Haurwitz, 1947; Rotunno, 1983; Dalu and Pielke, 1989; Li and Chao, 2016), as well as through simulations and observations (Yang, 1991; Atkison, 1995; Segal *et al.*, 1997; Steyn, 2003; Antonelli and Rotunno, 2007; Porson, Steyn and Schayes, 2007; Sills *et al.*, 2011; Cana, Grisolía-Santos and Hernández-Guerra, 2020). A sea breeze tends to form locally whenever a temperature difference between the air over land and that over sea persists (Musk, 1979), and it is strongly favoured when the background synoptic forcing is quiescent. Multiple physical factors affect the sea breeze evolution and the way it interacts with atmospheric boundary layer turbulence including: (i) the strength and directionality of the synoptic pressure forcing ($M_g$ and $α$, time-varying sometimes), (ii) the magnitude of the temperature contrast (Hadfield, Cotton and Pielke, 1992), (iii) the diurnal land surface temperature pattern (Weiming Sha, Kawamura and Ueda, 1991; Yoshikado, 1992; Miller *et al.*, 2003), (iv) initial stability conditions (Delage and Taylor, 1970), (v) relative surface roughness difference between the two surfaces (Finnigan, Shaw and Patton, 2009; Chen *et al.*, 2015; Jiang *et al.*, 2017), (vi) the Coriolis force (Antonelli and Rotunno, 2007), (vii) top and along-shore boundary conditions, and (viii) the coastline curvature (indentations in the shore line contour) and coastal orography (Collier, 2006). Therefore, the structure of an established LSB at any instant of time is highly sensitive to the non-linear interplay between these physical drivers and parameters, as well as to their history and that of the LSB itself. Hence, more observations and simulations are indispensable to advance our theoretical understanding of these circulations in this very extensive parameter space.

A specific challenge is to understand the relative importance of the physical mechanisms that drive LSB dynamics, as well as the fine details of the flow and transport patterns that emerge and influence coastal environments and populations. Despite the extensive parameter space, previous research has primarily focused on the temperature/heat flux contrast and how its magnitude modulates the flow, with other key physical drivers given little attention. The bulk of prior work pertains to idealized simulations of steady-state LSBs, often with no synoptic wind (Delage and Taylor, 1970; Hadfield, Cotton and Pielke, 1991), or to observational and numerical case studies that do not cover the full synoptic wind speed and direction parameter range (Estoque, 1962; Bechtold, Pinty and Mascart, 1991; Hadfield, Cotton and Pielke, 1992; Banta, Olivier and Levinson, 1993; Corner and McKendry, 1993; Atkins and Wakimoto, 1997; Gilliam, Raman and Niyogi, 2004; Drobinski *et al.*, 2006; Azorin-Molina and Chen, 2009; Qian, Epifanio and Zhang, 2009). To bridge this gap, this paper aims to explore a broader set of conditions (though the full parameter space remains too extensive for a single study). Specifically, we seek to answer the following question: (Q1) How do the synoptic variables ($M_g$ and $α$) modulate the fine-scale dynamics and structures of the LSBs? All other physical factors are here kept unchanged, and for simplicity the coastline is assumed to be a straight line. The complexity of the LSBs that emerge from that analysis, and their interaction with the surface and the shore (coastal zones), then motivate the second question: (Q2) How do the mass and thermal exchange fluxes at the shoreline scale with the external parameters $M_g$ and $W_*$ for all $α$'s? Finally, we reduce the results into a small set of archetypical circulation regimes that are visually identified at first, which then motivates the last question: (Q3) What are the most effective criteria to objectively classify the resultant LSBs in the ($Ri^{-1}$, $α$) regime space diagram? $Ri$ here is a heterogeneity Richardson number that we will define later.

The analyses rely on a set of large eddy simulations where we introduce the effects of $M_g$, oriented along a set of $α$ values, to a flow that is initially a quasi-steady LSB resulting from the case with quiescent background ($M_g = 0$). In section 2, a dimensional analysis framework for understanding this problem is presented. In section 3, the numerical experiments design is detailed. The thermal circulation flow structures are then discerned, explained, and visually categorized in section 4 (answering Q1). In section 5, land and sea surface kinematic fluxes and the characteristic sea breeze depth are investigated. Scaling of the net volumetric flux and modelling of the net advective heat flux at the shoreline are also explored (answering Q2). In section 6, two criteria are proposed to objectively demarcate these LSBs (answering Q3) with the aid of the *k*-means clustering algorithm (Lloyd, 1982) and the sensitivity of the LSBs dynamics is investigated. Finally, conclusions and implications of this work are drawn in section 7.

## 2    Dimensional analysis

Coastal dynamics can be captured by a set of dimensionless groups that can be constructed based on Buckingham's Pi theorem. In this paper, we attempt such an approach for a relatively simple setup that nonetheless is more realistic than studies of LSBs without synoptic forcing. The pertinent dimensional external input parameters to our problem involve three length scales of dimension (L): (i) $z_{0,L}$ = 0.002 (m), the land surface roughness length, (ii) $z_{0,S}$ = 0.002 (m), the sea surface roughness length (here not a function of wind), and (iii) $z_i$ = 1600 (m), the atmospheric boundary layer (ABL) height. In addition, two velocity scales of dimension (LT$^{-1}$) are of relevance: (i) $M_g = [U_g^2 + V_g^2]^{1/2}$ the geostrophic wind magnitude, and (ii) $W_* = [g\beta z_i(\theta_L - \theta_S)]^{1/2}$, a convective buoyant velocity scale where $g$ = 9.81 (m s$^{-2}$) is the gravitational acceleration, $\beta = 1/\theta_r$ (K$^{-1}$) the volumetric expansion coefficient of air, $\theta_r$ = 300 (K) a reference temperature in the Boussinesq approximation sense, and $\theta_L - \theta_S$ = 288.15 – 278.15=10 (K) the surface temperature difference between land and sea that is here kept constant. Finally, the flow physics are also influenced by one time scale of dimension (T) arising from the inertial period $2\pi/f_c \approx$ 12.5 h where $f_c$ = 1.394×10$^{-4}$ Hz is the Coriolis parameter, and by one non-dimensional variable, which is the alignment angle $\alpha$ between the shore and the geostrophic velocity vector.

Therefore, any output of the numerical experiments, with n=7 input dimensional parameters and k=2 independent dimensions (L, T), can be also non-dimensionalized and expressed in terms of n−k=5 dimensionless groups. Two dimensionless parameters are related to the introduced length scales $\Pi_1 = z_{0,L}/z_{0,S}$ (surface roughness contrast expressed as inner scales ratio) and $\Pi_2 = z_i/z_{0,S}$ (outer to inner scales ratio, commensurate with a Reynolds number of the problem). Another dimensionless group is associated with the velocity scales $\Pi_3 = Ri^{-1} = M_g^2/[g\beta z_i(\theta_L - \theta_S)] = M_g^2/W_*^2$, which is an inverse bulk Richardson number that relates inertia to buoyancy. An inverse convective Rossby number also arises, comparing the convective/buoyant and inertial time scales $\Pi_4 = (z_i/W_*)/(2\pi/f_c)$. The last group that remains is then $\Pi_5 = \alpha$. As stated earlier, the conducted simulations only consider the effects of the synoptic variables ($M_g$ and $\alpha$), and therefore $\Pi_3 = Ri^{-1}$ and $\Pi_5 = \alpha$ are the only inputs that we vary. $\Pi_1$ is set to 1 to isolate the impact of the thermal torque (due to temperature differences) from that of the stress torque (that is mainly controlled by roughness differences) (Bou-Zeid *et al.*, 2020). In addition, $\Pi_2 = 8\times10^5$ is fixed to a high value ensuring a high Reynolds number and large separation of turbulent to friction scales, while $\Pi_4 = 1.55\times10^{-3}$ is kept constant to remove any effect of variability in the role of the Coriolis force.

## 3 Large eddy simulations

### 3.1 Governing equations

In LES, the most dynamically energetic scales attributed to the energy-containing motions (i.e., scales that are most affected by the flow geometry and exhibiting a non-universal character) are directly resolved. On the other hand, eddies with scales smaller than a given cut-off (filter or grid) are parameterized. The used LES code solves the spatially-filtered incompressible mass continuity and Navier-Stokes momentum equations using the Boussinesq approximation for the mean state, in addition to the advection-diffusion equation for the potential temperature, which are given respectively as follows:

$$\frac{\partial \tilde{u}_i}{\partial x_i} = 0 \tag{1}$$

$$\frac{\partial \tilde{u}_i}{\partial t} + \tilde{u}_j\left(\frac{\partial \tilde{u}_i}{\partial x_j} - \frac{\partial \tilde{u}_j}{\partial x_i}\right) = -\frac{1}{\rho}\frac{\partial \tilde{p}^*}{\partial x_i} - \frac{\partial \tau_{ij}}{\partial x_j} - g\frac{\tilde{\theta}'}{\theta_r}\delta_{i3} + f_c(U_g - \tilde{u}_1)\delta_{i2} - f_c(V_g - \tilde{u}_2)\delta_{i1} \tag{2}$$

$$\frac{\partial \tilde{\theta}}{\partial t} + \tilde{u}_j\frac{\partial \tilde{\theta}}{\partial x_j} = -\frac{\partial \pi_j}{\partial x_j} \tag{3}$$

The resolved momentum equation is expressed in rotational form to improve mass and kinetic energy conservation (Orszag and Pao, 1975), and the tilde (~) represents filtered quantities (omitted throughout the rest of paper for simplicity since we only deal with filtered LES outputs); $x_i$ (or $X_i$) is the position vector; $\tilde{u}_i$ (or $U_i$) is the resolved velocity vector ($i$ and $j$ span the three directions 1 for $X$ (across shore), 2 for $Y$ (along shore), and 3 for $Z$ (vertical)); $t$ is time; $\rho$ is the density of air; $p^*$ is a modified perturbation pressure that include the resolved and unresolved turbulent kinetic energy; $\tau_{ij} = \widetilde{u_i u_j} - \tilde{u}_i \tilde{u}_j$ is the anisotropic part of the sub-grid scale (SGS) stress tensor; $\theta$ is the potential temperature with its reference value $\theta_r$; $\theta'$ is the deviation of the local $\theta$ from its horizontal planar average in the LES; $\delta_{ij}$ is the Kronecker delta; and $\pi_j = \widetilde{\theta u_j} - \tilde{\theta}\tilde{u}_j$ is the SGS heat flux vector. The SGS stress and heat flux are parameterized using a scale-dependent Lagrangian dynamic model that was extensively validated (Bou-Zeid, Meneveau and Parlange, 2005; Kumar *et al.*, 2006; Huang and Bou-Zeid, 2013), with a constant SGS Prandtl number of 0.4. The computational grid is a uniform structured mesh, staggered in the vertical direction to compute vertical derivatives using second-order centred differences. Horizontal derivatives are computed spectrally, and thus the domain must be periodic in the horizontal directions, but we modify it to impose inflow in the *X* direction, as clarified in the next subsection. An Adams-Bashforth second-order explicit time advancement scheme is used.

*3.2 Suite of simulations*

Sea breeze dynamics result in a circulation cell that manifest over a wide range of spatial scales, up to the depth of the atmospheric boundary layer depth (ABL). This could be aided or suppressed by the large-scale synoptic background. Therefore, it is useful to discriminate between two naturally occurring scenarios:

1. Null to weak synoptic forcing scenarios ($M_g$=0), mimicking coastal dynamical flows that are purely driven by thermal forcing with no imposed pressure gradients,
2. Medium to strong synoptic forcing scenario ($M_g \neq 0$), where the flow dynamics are driven by the interaction between the surface thermal forcing and the imposed background pressure gradients.

We designed LES cases to cover both scenarios, by varying the synoptic parameters ($M_g$ and $\alpha$) while keeping the temperature contrast constant. The designed simulations, illustrated in Fig. 1, span thirty cases with all combinations of {$M_g$ (m/s): 0.4, 0.8, 1.2, 1.6, 2 and $\alpha$: 0°, 45°, 90°, 180°, 225°, 270°}. The domain size is $L_x$ = 80 km × $L_y$ = 5 km × $L_z = z_i$ = 1.6 km. The corresponding baseline number of grid points ($N_x$, $N_y$, $N_z$) is 384×24×64. As indicated by (Atkison, 1995), some large length scales along the *Y* direction might exist as a result of the Kelvin–Helmholtz billows in the vicinity of the sea-breeze front structure, but the scale of these billows comprises a fraction of the gravity current depth (here thus limited to be $< z_i$) and hence lies well within the 5-km domain selected for the $Y$ direction. The grid resolution ($dX = dY = 208$ m, $dZ = 25$m) is course relative to typical ABL LES studies, constrained by the requirement to have a very large domain to resolve the extensive circulations in the cross-shore direction, and the need for many simulations to span the parameter space. The simulations thus represent a very large eddy simulation of the problem, where a significant fraction of the production range may not be resolved (Pope, 2001), but the analyses focus primarily on the mean flow that should still be captured well.

To ensure the robustness of the conclusions, two types of grid sensitivity analysis are conducted: one for the null synoptic forcing scenario ($M_g$=0) case and another for the ($M_g$=1.2, $\alpha$=180°) case. The reason $\alpha$=180° is chosen for this sensitivity analysis is that, as we will show later, this direction favours the development of a very stable boundary layer over the sea patch that might be quite sensitive to grid resolution. The first sensitivity test includes increasing the original grid resolution by a factor of 1.5 where $N_x$, $N_y$, $N_z$ become 576×36×96, while keeping the size of the computational domain fixed. The second test doubles $N_y$ (to 48) and $L_y$ (to 10 km) only (since in the baseline it is $\ll L_x$). The results indicate very similar circulation features and qualitative conclusions (refer to Figs. S1 to S4 and tables S1 to S5 in the supporting information (SI) section 1). However, complete grid convergence was not attained, and some of the quantitative results (especially second order moments i.e., surface heat fluxes) might vary moderately if finer grids are achievable in future studies. Most importantly, the errors in shore fluxes that we will rely on later are consistently small ($\ll$1%) for both sensitivity tests (and for the analysed statistics over different averaging periods) and will thus not affect the conclusions drawn in the modelling part and the scaling analysis (as we will show in subsections 5.3 and 5.4).

*3.3   Boundary and initial conditions*

For both scenarios (null or positive $M_g$), the boundary conditions are set periodic in the $Y$ direction for velocities and temperature in all simulations to mimic an infinite coastline. A stress-free impermeable lid boundary condition at top of the domain is imposed for the velocities because any resultant LSB might be sensitive to the free tropospheric profiles and inversion strength (Cioni and Hohenegger, 2018), and in this paper we want to avoid adding yet another parameter to the investigation (our setup mimics a very strong inversion). The surface temperatures of the two patches are imposed as $\theta_S$ = 278.15 K and $\theta_L$ = 288.15 K, and surface stresses and heat fluxes are calculated using a local equilibrium wall-model based on a log-law with Monin-Obukhov stability correction (Bou-Zeid, Meneveau and Parlange, 2005; Ghannam and Bou-Zeid, 2021). With the above boundary conditions in mind, we implemented a numerical setup that is as faithful as possible to the physics in both scenarios. Almost all previous simulations of the land-sea breeze use open along-shore boundaries (Orlanski, 1976; Miller and Thorpe, 1981) that pose little resistance to the flow leaving the domain (unless the domain is made very wide, at the expense of resolution), but real sea breezes act more like gravity currents into a quiescent fluid that thicken and are slowed by that fluid. Also, the open along-shore boundaries cannot be used in the second scenario with the synoptic pressure driven flow where an inflow is needed on one of the boundaries.

With these physical constraints in mind, we (i) implemented two equal buffer regions on the streamwise boundaries of the domain to act as smooth transitions of the streamwise flow (since the numerical code is pseudo-spectral in $X$), and (ii) generated realistic upstream inflow for a range of geostrophic wind conditions (for all $\alpha$ values) when the synoptic forcing is significant. As sketched in Fig. 2, the buffer regions are symmetric with respect to the periodicity interface $X=0$, and the length of the buffer area ($L_{Buffer}$) on each side is 1/16 of $L_x$ ($\approx$ 5 km). This width was found most suited to minimize the Gibbs phenomenon and wave reflection. This setup is broadly similar to the one used recently by Sullivan et al. (2021) for simulating marine boundary layer development over heterogeneous sea surface temperatures, a very similar problem to the one we study here (although they use a fringe region implementation for the buffer zone). Also note that, on either side of the domain in $X$, we exclude a region twice the buffer length from analysis (demarked by the dashed lines in Fig. 2) to ensure the streamwise boundary conditions impact on the analyses is minimized.

The $X$ boundary and initial conditions for velocities and temperature are treated differently in the two scenarios (with and without synoptic wind):

i) Null synoptic forcing scenario ($M_g=0$):

A no-slip boundary condition (NS-BC: $U=V=W=0$) is imposed at the end of the domain in the $X$ direction ($X=0$ and $X=L_x$). As sketched in the top left panel of Fig. 2, two equal buffer regions are set relative to this NS-BC consisting of $Y$-$Z$ planes that, at each time step, interpolate/recycle (tangent hyperbolically) the solutions of velocities and temperature from at those computed at the edge of the analysis domain to the imposed/desired wall values (Spalart, 1988; Lund, Wu and Squires, 1998). The desired temperature at the side wall ($X=0$) is taken as the average between the Y-Z plane at ($X=L_x-L_{Buffer}$) and Y-Z plane at ($X=L_{Buffer}$). In the buffer zone, the surface temperature is then also tangent-hyperbolically interpolated to an average temperature is $\theta_{avg} = (\theta_L+\theta_S)/2$ at the periodicity interface ($X=0$). The air temperature is uniformly initialized to the same average temperature, with $\theta_{init} = \theta_{avg}$ (an educated guess though the final temperature over long simulations would be closer to the hotter patch one), and therefore this initialization renders the sea stable and the land unstable. This is not a unique initialization, and we did find some sensitivity to the initial conditions in our simulations, mostly when we initialized with $\theta_{init}=\theta_L$, which rendered the sea strongly stable and the land neutral and needed a very long time to break the established stability over the sea. The case with $\theta_{init}=\theta_S$ produced similar results as the adopted $\theta_{init}=\theta_{avg}$. The analysed results here refer to the last half inertial period ($0.5\times\tau_f=6.25$ h) out of the total simulation time ($T=5\times\tau_f=5\times2\pi/f_c=62.5$ h) to focus on quasi-steady state dynamics, as illustrated in the middle panel of Fig. 2.

ii) Non-zero synoptic forcing scenario ($M_g\neq 0$):

The inflows for velocities and temperature are generated in precursor simulations with periodic boundary conditions in both $X$ and $Y$ directions. The imposed surface temperature (and thermal roughness lengths) in the precursor simulations is set equal to the value of the sea or land over the inlet patch in the main simulations to represent an infinitely homogeneous upstream fetch. The air temperature is uniformly initialized with $\theta_{init}=\theta_S$ for sea precursors and $\theta_{init}=\theta_L$ for land precursors (both neutral simulations). Saved inflows refer to the last quarter inertial period ($0.25\times\tau_f=3.125$ h) out of the total precursor simulations time ($T=4.25\times\tau_f=53.125$ h), and these inflows are recycled every $0.25\times\tau_f$ in the $M_g\neq0$ simulations (see middle subplot of Fig. 2). Since the internal dynamics of the main domain are evolving, this recycling will not produce identical flows repeated every $0.25\times\tau_f$ in the main domain. Again, the analyses consider the statistics in the last $0.5\times\tau_f$. The dynamics ($U$, $V$, and $W$) are interpolated in the buffer zone exactly as described above for the $M_g=0$ case, but here the interpolation is to/from the imposed precursor inflow planes that are generated for each $M_g$ and $\alpha$ value (top right of Fig. 2). We also tested the fringe region technique (generally similar to what we use here), proposed by (Stevens, Graham and Meneveau, 2014) and (Munters, Meneveau and Meyers, 2016) to circumvent the periodic numerical boundary condition, but the simulations produced almost identical results. The air temperature fields are only interpolated in the buffer region over the patch opposing the inflow patch. The surface temperature is also interpolated only in that buffer zone, tangent hyperbolically, to the inlet patch surface temperature $\theta_{X=Lx}=\theta_{inlet}$ at $X=0$.

It is important to underline that air temperature and velocities, for $M_g \neq 0$ simulations, are initialized with the resultant fields of the null synoptic forcing scenario ($M_g=0$). This mimics a sudden change in the pressure gradient from an initial state where it is negligible. It is a realistic scenario that for example mimics a sudden onset of a driving synoptic gradient, e.g., near fronts... Furthermore, we continue to impose the same synoptic pressure gradient (as a geostrophic wind) used to generate the inflow in the main simulation domain, and thus we do not rely solely on the inflow inertia to represent the synoptic flow. A schematic diagram of the three-dimensional domain and the boundary conditions is given in the bottom subplot of Fig. 2.

## 4 Thermal circulation structures

The fine-scale dynamics and bulk structure of the sea breeze are not yet fully understood (Crosman and Horel, 2010b). We thus begin by analysing the features of the simulated LSBs, and how they are jointly modulated by $M_g$ and $\alpha$. Figs. 3 and 4 depict pseudocolour plots of the along-shore and time averaged (henceforth denoted by angled brackets) streamwise $\langle U \rangle_{y,t}$ through an X-Z slice of the analysis domain [$X$=10 km-70 km]. The series of subplots in each figure correspond to the same $\alpha$ with increasing $M_g$. Everywhere in this paper, the streamlines are drawn in magenta colour when the surface wind, absent a LSB, would blow from sea towards land ($\alpha=0°$, $\alpha=45°$, and $\alpha=270°$) and in black colour for the opposite direction ($\alpha=180°$, $\alpha=225°$, and $\alpha=90°$). The very first subplot in these figures — and all similar figures in the paper — refers to the baseline null synoptic forcing scenario ($M_g=0$). Similarly, Figs. 5 and 6 show pseudocolour plots of the temperature $\langle \theta \rangle_{y,t}$. The selected colour bar ranges for $\langle U \rangle_{y,t}$ (m/s) and $\langle \theta \rangle_{y,t}$ (K) are kept unchanged in all figures to provide a basis for a clear comparison. Four regimes emerge in these figures, as detailed in the following 4 subsections.

*4.1 Canonical LSB*

The first subplot ($M_g$=0) in Fig. 3 exemplifies a quasi-steady picture of the canonical deep thermal circulation that results from a steady thermal contrast in the absence of synoptic pressure forcing. Near the surface, hot/cold air parcels rise/descend over the land/sea due to buoyancy, thus creating a pressure gradient that drives cold air from sea towards land to replenish the uplifted air (creating the sea breeze). At higher levels, the pressure gradient reverses and so does the airflow direction; therefore, the counter-clockwise circulation closes by continuity (advection of hot air from land towards sea at higher elevations). Near the surface, $\langle U \rangle_{y,t}$ intensifies gradually with respect to $X$, and it reaches around 4 m/s on land (note that a sea breeze front is not observed in this steady simulation, we do observe it in transient diurnal simulations we are conducting at present). The stalled air region where the flow reverses is demarcated by the white area when $\langle U \rangle_{y,t}$ changes sign. This region forms a wedge-like shape towards the mid-height of the domain with $M_g = 0$, and its width increases with respect to $X$.

This picture is consistent with the thickness of the internal thermal boundary layer $\delta_{th}(X)$ as shown in the corresponding (upper left) subplot of Fig. 5. This figure also displays the canonical temperature field anticipated for an LSB. Generally, near the surface, there is a tendency for enhanced mixing over the land patch because the cooler sea breeze renders the land highly unstable ($d\langle \theta \rangle_{y,t}/dz|_s < 0$). Conversely, the return of hot air from land towards sea at higher altitudes and its subsequent subsidence renders the sea stable ($d\langle \theta \rangle_{y,t}/dz|_s > 0$), thus suppressing mixing and favouring the growth of an internal stable boundary layer.

This canonical LSB pictured also holds for all simulations with a geostrophic wind parallel to the shore (Figs. S9 and S11 of SI section 2). All the subplots of these figures depict slightly modified versions of the $M_g$=0 canonical LSB, hinting at the insignificant role of $V_g$ in changing the bulk LSB flow structure. However, the corresponding $\langle \theta \rangle_{y,t}$ temperature fields in Figs. S10 and S12 are much warmer compared to $M_g$=0. This is probably due to the enhanced near-surface turbulence (stronger fluxes) and strengthened surface wind and heat fluxes, especially over land, created by the stronger along-shore velocity component. As $M_g$ increases, the $X$ component due to Ekman rotation increases too but in opposite directions for $\alpha$=90° (land to sea) and $\alpha$=270° (sea to land). Hence, along-shore cases $\alpha$=90°/270° damp/boost the LSB strength marginally but do not change its bulk structure. (Estoque, 1962) also noted that along-shore cases have minimal effect on changing the LSB, but that the nature of the differences between both directions is less evident.

*4.2 Land-driven LSB*

A different regime that arises when the synoptic wind blows from land to sea, the land-driven LSB, features the development of a persistent LSB that is however much shallower than the canonical LSB. Another feature of this LSB, inherent to land-driven cases, is that it is long-lived and resistant to the increased synoptic pressure forcing (compared to the sea driven regime we will discuss next), and thus it endures even at the highest $M_g$ we simulate here. In the subplots of Fig. 3 ($\alpha=180°$; $M_g=U_g \geq 0.4$ m/s), the synoptic inflow from the land impedes the sea breeze of the already established canonical LSB near the surface, while boosting the flow of the return air above. The synoptic wind also depletes the energy of the circulation because it renders the land more neutral near the surface (since inflow is from the land here). In addition, the synoptic flow aids the return and downward sweep of hot air from land, thus strengthening the stability over the sea patch.

At $M_g=U_g=0.4$ m/s, the apparently marginal perturbation in the synoptic forcing has a strong impact on the canonical LSB (the initial condition of the simulation), but its structure is overall maintained. As seen in this subplot, the sea breeze is mildly attenuated and thinned compared to the canonical LSB whereby $\langle U \rangle_{y,t}$ only reaches 3 m/s. This also significantly impacts the corresponding subplot of $\langle \theta \rangle_{y,t}$ in Fig. 5 as the synoptic wind here changes the entire range of $\langle \theta \rangle_{y,t}$ when compared to the canonical LSB. The neutral $\langle \theta \rangle_{y,t}$ land inflow profile persists inland up to ~ 10 km from the coast, replacing the more unstable regime prevalent for $M_g = 0$. In addition, the boosted hot return air — aided by land inflow — helps in generating a very stable shallow thermal boundary layer of almost 10 K difference between the sea surface and the air aloft, but the LSB maintains its canonical type.

However, as $M_g$ increases further for $\alpha=180°$, the LSB structure begins to transition towards what we here refer to as a land-driven LSB, and this transition is highly sensitive to the interaction between buoyancy, stability, and synoptic forcing. In the subplot of Fig. 3 with $U_g=0.8$ m/s, the large LSB is obliterated and only a very shallow but strong sea breeze is seen, with $\langle U \rangle_{y,t} \approx 3$ m/s. This results from the enhanced stability over the sea patch near the surface, and the sea breeze remains stronger than the return air in the upper part of the circulation where $\langle U \rangle_{y,t} \approx U_g = 0.8$ m/s. The corresponding subplot of $\langle \theta \rangle_{y,t}$ in Fig. 5 exhibits a neutral temperature profile at $\langle \theta \rangle_{y,t} \sim \langle \theta \rangle_{y,t}|_{\text{land}}$ extending over much of the land area. The hotter (relative to $M_g = 0$ case) return air helps in generating a more stable shallow thermal boundary layer over the sea patch. Note how the temperature colour range amplifies across the whole analysis domain in the subsequent subplots as $M_g$ increases, and the maximum air temperature approaches the inflow land temperature (dark red strip).

The shift towards land-driven LSB amplifies in the rest of the subplots of Fig. 5 ($M_g$=1.2, 1.6 and 2 m/s); the flow structures resemble the $M_g$=0.8 case, except that the sea breeze weakens further. Moreover the onshore penetration distance of the sea breeze front is impeded by the incoming inflow to around 8 km from the shore as in the subplot for $M_g$=2 (bottom right), a severe reduction even compared to the $M_g = 1.2$ where it spans almost the whole land patch. Such significant penetration of the sea breeze under strong synoptic forcing has been observed even for geostrophic flows of $M_g \approx$ 4-5 m/s (Arritt, 1993a; Tijm, Van Delden and Holtslag, 1999; Porson, Steyn and Schayes, 2007).

The corresponding subplots of $\langle \theta \rangle_{y,t}$ in Fig. 5 show almost the same features for all $M_g \geq 0.8$, and therefore the intensely stable thermal boundary over the sea persists in all scenarios. The increased pressure forcing along $\alpha$=180°, up to the limit of $M_g = U_g =$ 2 m/s we reach in this study, does not fully collapse the established internal thermal boundary layer over the sea patch (a shallow sea to land breeze persists) because stronger incoming land inflows correlates with stronger stabilities over the sea patch. The resultant LSBs lose symmetry relative to the shoreline, as also observed by (Finkele *et al.*, 1995). Furthermore, as $M_g$ increases, shear instabilities (that might result due to the rapid change in stratification) near the coastline hinder SB inland penetration as simulated by (Grisogono, Ström and Tjernström, 1998).

In the subplots of (Figs. S7 and S8 of SI section 2, $\alpha$=225°; $U_g=V_g=-M_g/2^{1/2}$), the direction of the pressure forcing here is still impeding the sea breeze of the already established canonical LSB near the surface and boosting the flow of the return air above. However, but this interaction is weakened compared to the respective $\alpha$=180° cases. This again indicates that $U_g$ is the main component responsible for the modulation of the whole LSB.

### 4.3 Sea-driven LSB

Sea-driven LSBs emerge when the geostrophic wind blows from the sea. In our setup, we observe the formation of two circulation cells, where a new clockwise circulation cell emerges above the old counter-clockwise canonical one. However, this new cell (as will be depicted later) is short-lived and weakly resistant to increased synoptic pressure forcing, and thus it is prone to merging back with the old canonical cell or eventually getting advected. The upper cell also will likely depend on the top boundary condition of the domain, so it is not a focus of our discussions.

The inflow here consists of neutral Y-Z slices generated over a periodic sea domain, with flow almost perpendicular to the shore. In the subplot of Fig. 4 with $α=0°$ and $M_g=0.4$ m/s, we can see that the synoptic wind aids the sea breeze of the already established canonical LSB near the surface and suppresses the return flow above. At weak $M_g$, synoptic winds from the sea thus bolster the energetics of the circulation. This is because the land becomes more unstable near the surface, and the flow now inhibits the return of hot air from land, thus curbing the growth of the internal stable thermal boundary layer over the sea patch. In addition, the existing canonical LSB that was driven solely by buoyancy, is now also energized by the introduced mechanical energy resulting in what we refer to as a sea-driven LSB. The strength of the sea breeze now increases to a new maximum $\langle U \rangle_{y,t} > 4$ m/s. Near the top at around $Z=1300$ m, the synoptic inflow squeezes the remnants of the LSB return branch to lower altitudes, forcing it to accelerate to a new maximum of $\langle U \rangle_{y,t} \approx -4$ m/s (dark blue area). While these dynamics will be sensitive to the exact inversion strength, domain top BCs, and other details, they do indicate a shallower sea breeze pattern that should be robust to the numerical and physical details. Analysing the corresponding subplot of the temperature field in Fig. 6 demonstrates the persistent neutral profile of the advected inflow over the sea, which then slowly warms as it passes over land towards the right end of the active domain. Similarly, the stable thermal boundary layer that forms below the return branch is now attenuated and lofted to higher altitudes. When $M_g$ increase to 0.8 m/s in Fig. 4 and Fig. 6, the velocity and temperature fields remain almost identical to the $M_g=0.4$ m/s case, except that the SB intensifies further. Note how the temperature range narrows across the whole analysis domain in these subplots, and the maximum air temperature approaches the inflow sea temperature.

As $M_g$ increases above 0.8 for $α=0°$, a sudden regime shift occurs that we will detail in the next subsection. However, a sea driven LSB can still be noted for $α=45°$ at $M_g=0.8$ and 1.2 m/s (see Figs. S5 and S6 of SI section 2, $α=45°$; $U_g=V_g=M_g/2^{1/2}$). The direction of the pressure forcing and inflow here is still aiding the sea breeze near the surface and suppressing the return air above, but this interaction is weakened compared to $α=0°$ with corresponding $M_g$. This is again due again to the dominance of $U_g$, over $V_g$, in driving the whole LSB.

*4.4 Advected LSB*

Advected LSBs display an almost unidirectional flow across the whole ABL domain, and therefore such flows do not feature any circulation cell, but they might inherit some weak remnants of freshly eradicated LSBs.

In the subplots of Fig. 4 ($\alpha=0°$; $M_g=U_g=1.2$ and 1.6 m/s), the flow is almost completely pressure driven, with minimal return air. The stalled air region and return flow are confined to a small zone in the upper right most part of the domain. Furthermore, the return branch is severely impeded by the incoming inflow to around $X=30$ km, as shown in the $M_g=1.6$ subplot; however, it spans almost the whole sea patch in the $M_g=1.2$ subplot. The corresponding subplots of $\langle\theta\rangle_{y,t}$ in Fig. 6 demonstrate almost the same $\langle\theta\rangle_{y,t}$ profile as the inflow (dark blue strip, stronger advection of the neutral sea profile in the whole analysis domain).

In the subplot of Fig. 4 ($\alpha=0°$; $M_g=U_g=2$ m/s), the flow is purely pressure driven, with no return air. The stalled air region and circulation cell are obliterated, with only small inland traces of the initial domain air. Here, the thermally generated baroclinic pressure gradient is cancelled by the synoptic pressure gradient (Crosman and Horel, 2010b). The corresponding subplot of $\langle\theta\rangle_{y,t}$ in Fig. 6 displays a persistence of the $\langle\theta\rangle_{y,t}$ inflow profile over the whole sea patch and protruding inland as well to reach $X\approx 45$ km, with mild warming subsequently. In the literature, there is no consensus on a critical value of $M_g$ at which an LSB does not form, as there are many factors that affect this threshold (i.e. body dimension of the sea-land configuration, thermal contrast strength, and all the factors listed in the introduction), instead there are wide reported ranges of this critical $M_g$ (Arritt, 1993a; Crosman and Horel, 2010b).

In a nutshell, the pronounced increase of the pressure forcing in the $\alpha=0°$ configuration is more weakly resisted by the thermal circulation than at $\alpha=180°$. For the $\alpha=0°$ scenarios, the circulation is pushed onto the land by the incoming sea inflows that render the land patch highly unstable near the surface, which enhances mixing over the entire patch and destroys any circulation. Thus, as $M_g$ increases over the simulated range [0 to 2 m/s] for $\alpha=0°$, the sea remains neutral, and the land goes from mildly to extremely unstable. Conversely, for $\alpha=180°$ where the circulation is pushed towards the sea, the sea remains stable near the surface, while the land goes from extremely unstable to neutral again. There are thus only weak convective plumes near the surface to completely mix the circulation. Similarly, the subplots of Fig. S5 ($\alpha=45°$; $M_g=1.6$ and 2 m/s), which also show an advected LSB type, represent attenuated/delayed versions of the corresponding subplots of Fig. 4 ($\alpha$=0°; $M_g$=1.6 and 2 m/s). A quick comparison between the subplot of Fig. 4 ($\alpha$=0°; $M_g$=1.2 m/s) and that in Fig. S5 ($\alpha$=45°; $M_g$=1.2 m/s) shows how $\alpha$ can shift the regime from an advected LSB to a persistent sea-driven LSB, for the same geostrophic wind speed.

## 5 Surface and shore exchanges

### 5.1 Surface heat flux

Given the importance of the changes in surface heat flux in explaining the modulation of the LSB as geostrophic wind speed increases, we plot in Fig. 7 the averaged surface heat flux $\langle w'\theta'\rangle_{y,t}$ over the two analysis patches. The sea fluxes are then also averaged over $X$=10-40 km, while land fluxes are averaged over $X$=40-70 km. The first column of circles in each subplot refers to the canonical scenario $M_g$=0, which represents both the baseline scenario and the initial conditions to the simulation results of the consecutive columns. In the left subplot of Fig. 7, the sea surface heat flux of the $M_g$=0 scenario is slightly stable with $\langle w'\theta'_{Sea}\rangle_{y,t} \approx -0.003$ K m/s ($-3.6$ W/m²). The rows corresponding to $\alpha$=0° and 45° show that any increase in $M_g$ drives the sea to a neutral state. As discussed above, this is due to the enhanced advection of neutral sea inflow planes over the sea and then towards the land. The initial sea stabilizing heat flux is almost preserved along $\alpha$=270°, with a slight increase at $M_g$=0.8 m/s followed by a return to the initial flux values at higher $M_g$. The same insensitivity is noted for $\alpha$=90°, both cases pertaining to a pressure driven flow along the shore that does not impact the circulation significantly; however, by modifying wind speed near the surface, they do change the fluxes moderately.

On the other hand, the row corresponding to $\alpha$=180° demonstrates that the sea fluxes are extremely sensitive to any increase in $M_g$ in that geostrophic orientation. The stabilizing sea heat flux is rapidly enhanced with increased $M_g$. Here, $\langle w'\theta'_{Sea}\rangle_{y,t}$ reaches about $-0.013$ K m/s ($-15.6$ W/m²), almost one order of magnitude higher than for $M_g$=0. The increased flux peaks at $M_g$=1.6 m/s, and then it slightly decreases at $M_g$=2 m/s. The row corresponding to $\alpha$=225° shows a consistently delayed picture of the row corresponding to $\alpha$=180°, but, interestingly, the sea flux at ($\alpha$=225°, $M_g$=2 m/s) attains even higher value $\langle w'\theta'_{Sea}\rangle_{y,t} \approx -0.015$ K m/s ($-18$ W/m²) compared to the corresponding $\alpha$=180°.

In the right subplot of Fig. 7, the land surface heat flux of the $M_g$=0 scenario is unstable with $\langle w'\theta'_{Land}\rangle_{y,t}$ ~0.06 K m/s (72 W/m$^2$). The rows corresponding to $\alpha$=180° and 225° show that perturbations to the $M_g$=0 drives the land neutral. This is due to the enhanced advection of neutral land inflow planes from land towards sea (from the precursor, set here as neutral). Although the land in reality might have an unstable ABL, the land surface-air temperature contrast is expected to be much smaller than the land-sea temperature contrast. For the parallel cases of $\alpha$=90° and $\alpha$=270°, the land stability fluctuates marginally relative to the $M_g$=0 case. On the other hand, the $\alpha$=0° case exhibits an extreme sensitivity to any perturbation to $M_g$. The land's unstable regime is enhanced along this direction with increased $M_g$. As the sea breeze intensifies and flows over the hotter land, the absolute temperature gradient increases, a process known as sea breeze frontogenesis and defined by (Miller *et al.*, 2003)). Here, $\langle w'\theta'_{Land}\rangle_{y,t} \approx 0.11$ K m/s (132 W/m$^2$) is almost twice the magnitude of the initial flux. Similarly, the amplified unstable regime peaks at $M_g$=1.6 m/s, and then slightly decreases at $M_g$=2 m/s. The row corresponding to $\alpha$=45° again shows a delayed similarity to the row corresponding to $\alpha$=0°. The land heat flux at ($\alpha$=45°, $M_g$=2 m/s) reaches even higher intensity $\langle w'\theta'_{Land}\rangle_{y,t} \approx 0.12$ K m/s (132 W/m$^2$) compared to ($\alpha$=0°, $M_g$=2 m/s). A complementary analysis of the sea and land friction velocity was conducted, but since it has a minimal effect on the circulation and discussions, we present it in SI section 3 (Fig. S13).

*5.2 Sea breeze depth*

The sea breeze depth is limited by the boundary layer height at the shore when the capping inversion is strong (Simpson, 1969; Reible, Simpson and Linden, 1993), but it is often shallower and therefore it is important to study its behaviour with respect to changes in $M_g$ and $\alpha$. The $SB_{depth}$ is here defined as the height at which the $\langle U_{Shore}\rangle_{y,t}|_{X=40km}$ switches sign from positive (sea breeze) to negative (return air aloft). In Fig. 8, for $\alpha$=0°, the initial $SB_{depth}$ is around 812 m for the $M_g$=0 scenario. Then, as $M_g$ is increased, this introduced pressure forcing modulates the circulation non-monotonically due to the competition and interaction between buoyancy and shear. As a result, $SB_{depth}$ first decreases when the LSB is of the sea-driven type. However, as $M_g$ is further increased, the $SB_{depth}$ increases again when the sea breeze becomes of the advected LSB type. For the $M_g$=2 m/s, the thermal circulation is eliminated (at least at the shore) and the $SB_{depth}$ cannot be determined. This nonlinear $SB_{depth}$ behaviour is attributed to the changes of surface heat fluxes as described in the previous subsection and the resultant intensity of boundary layer convection inland because of the sea breeze (Garratt, 1990; Miller *et al.*, 2003). In fact, the effect of geostrophic winds lower than 4 m/s on $SB_{depth}$ remain a subject of disagreement in the literature (Crosman and Horel, 2010b). The results of this paper (where $M_g$ is less than 2 m/s), show that the $SB_{depth}$ is jointly affected by synoptic winds, mechanical turbulence, buoyancy, and frontogenesis, which may explain why previous studies remain divergent on its dynamics. For $α=45°$, the initial $SB_{depth}$ follows the same pattern as the ($α=0°$) cases but features shallower depths because $U_g$ is decreased. One difference is that for $M_g=2$ m/s, the advected thermal circulation persists and the $SB_{depth}$ attains 1219 m unlike the $α=0°$ where $SB_{depth}$ was indeterminate.

At $α=180°$ for land-driven LSB cases, the initial $SB_{depth}$ decays sharply and monotonically as $M_g$ is increased until it plateaus at a very small elevation of about 40 m. Similarly, for $α=225°$, the initial $SB_{depth}$ follows the same pattern as the $α=180°$ cases; however, the resulting $SB_{depth}$ is deeper at the same value of $M_g$ compared to $α=180°$. The $SB_{depth}$ for $α=225°$ plateaus around 75 m for $M_g=2$ m/s. For these two wind angles, the noticeable decrease in $SB_{depth}$ as $M_g$ is increased even mildly agrees with the general findings of (Arritt, 1993b; Zhong and Takle, 1993), and the observed plateau as $M_g$ is increased further in the very last increments agrees with the general conclusions of (Estoque, 1962). However, our study is the first to methodically link the $SB_{depth}$ unique behaviour to specific $M_g$ ranges and orientations. For $α=270°$ and $90°$, the $SB_{depth}$ fluctuates nondeterministically around the initial $SB_{depth}$ as $M_g$ is increased.

### 5.3  Volumetric flux at the shore

Fig. 9 shows four subplots of the shore volumetric flux ($Q_{shore}$) variation, for each $α$, with respect to $Ri^{1/2}$. The left panel of subplots corresponds to dimensional quantities of $Q_{shore}$, and the right one represents a normalized version $Q_{sc}$ (defined below). $Q_{shore}$ is defined as the net volumetric flux across a unit along-shore width, $\|y_n\|=1$ m (added merely for convenience in interpreting units), at the shore interface $X=40$ km ($Q_{shore} = Q_{shore}^{SL} - Q_{shore}^{LS}$). Hence a positive $Q_{shore}$ value implies that the sea breeze $Q_{shore}^{SL}$ is stronger than the return air aloft $Q_{shore}^{LS}$ (more fresh air is brought from sea towards land), and a negative $Q_{shore}$ value signifies the opposite. Thus,

$$Q_{shore} = Q_{shore}^{SL} - Q_{shore}^{LS} = \frac{\|y_n\|}{L_y} \int_0^{L_y} \int_{z_0}^{L_z} U_{shore} dz\, dy \tag{4}$$

$$Q_{sc} = \frac{Q_{shore}}{M_g L_z \|y_n\|} = \frac{1}{L_z L_y M_g} \int_0^{L_y} \int_{z_0}^{L_z} U_{shore} dz\, dy \tag{5}$$

For $\alpha=0°$ in the top left of Fig. 9, as $Ri^{1/2}$ is decreased implying strong $M_g$, $Q_{shore}$ increases steeply to reach around 3350 m³/s for $M_g= 2$ m/s (almost 4.5 times increase relative to $M_g=0.4$ m/s). This is predictable as stronger sea inflow is advected. Furthermore, the trend is quite similar along $\alpha=45°$, with weaker values compared to $\alpha=0°$ ($U_g$ is attenuated here). The same picture noted for $\alpha=0°$ and $\alpha=45°$ prevails for the symmetric directions $\alpha=180°$ and $\alpha=225°$, respectively. For $M_g=2$ m/s, $Q_{shore}$ reaches around $-3026$ m³/s and $-2062$ m³/s along $\alpha=180°$ and $\alpha=225°$, respectively, which are slightly lower fluxes than the value reported for $\alpha=0°$ and $\alpha=45°$. This difference is because the sea breeze is never eliminated near the surface along $\alpha=180°$ and $\alpha=225°$, even for $M_g=2$ m/s.

The trends with $\alpha=90°$ and $\alpha=270°$ in the bottom left of Fig. 9 are different, nonmonotonically decreasing with $Ri^{1/2}$. Recall that along these directions ($M_g=V_g$), the pressure forcing (irrespective of its strength) affects the canonical LSB only slightly, but never changes its canonical type. $Q_{shore}^{SL}$ remains stronger than $Q_{shore}^{LS}$ for all $M_g$ here (resulting in a consistent positive $Q_{shore}$). However, note the diminished values (which make these results more prone to numerical errors and boundary condition and buffer zone sensitivity) and ranges of variability for $Q_{shore}$ (one order of magnitude less than the reported values for the other $\alpha$'s in the top left subplot). $V_g$ here tends to preferably advect more mass along the periodic $Y$ direction, but a slight Ekman rotation that creates an $X$ component is responsible for the difference in the two trends of $\alpha=90°$ and $\alpha=270°$. As $Ri^{1/2}$ decreases, the cross shore ($X$) Ekman component increases too, but in opposite directions: for $\alpha=90°$ (land to sea) it weakens the flux while for $\alpha=270°$ (sea to land) it boosts the flux. This explains the mildly increasing trend for $\alpha=90°$ and the decreasing trend for $\alpha=270°$ with increasing $Ri^{1/2}$.

The right panel of subplots of Fig. 9 shows the normalized version ($Q_{sc}$). Postulating that $Q_{shore} \propto n M_g + m W_*$, a mixed velocity scale, results in $\frac{Q_{sc}}{cos(\alpha)} = \frac{Q_{shore}}{M_g\, cos(\alpha)} \propto \frac{n M_g + m W_*}{M_g\, cos(\alpha)} \propto n + m\, Ri^{\frac{1}{2}}$. A corollary of this scaling is that the external parameter $Ri^{1/2}$ and internal parameter $Q_{sc}$ can be related via $Ri^{1/2} = a_i \left(\frac{Q_{sc}}{cos(\alpha_i)}\right) + b_i$. The coefficient $a_i$ and $b_i$ will be treated as fitting parameters that can depend on wind angle, and thus are separately determined for the sea-to-land geostrophic wind directions, $\alpha_{1,2}=0°$ and $45°$, and the land-to-sea directions $\alpha_{4,5}=180°$ and $225°$. As for the along-shore directions, $\alpha_3=90°$ and $\alpha_6=270°$, the analysis cannot involve any $cos(\alpha)$ scaling as $\cos(\alpha_{3,6})$=0. While one can attempt to scale $Q_{shore}$ with $W_*$ alone for these cases, the circulation strength was shown to be sensitive to $M_g$ in the previous sections. Thus, we postulate $Ri^{1/2} = a_i Q_{sc} + b_i$.

While this analysis may not yield universal models that can be broadly used before further efforts to generalize them, it proves useful as a tool to elucidate the physics of the competition between buoyancy and pressure forcing, as we will illustrate. The black lines in Fig. 9 (right panels) correspond to this scaling, and the fitting parameters are summarized in table 1. The reported $R^2$ values ($R^2 > 0.732$ for all $\alpha$'s and $R^2 \approx 1$ for $\alpha_3$=90° and $\alpha_6$=270°) show the high skill of the suggested models in describing the LES results. This is clearly discerned in the collapse of data in the top right subplot for the opposite across-shore directions $\alpha_{1,2}$ and $\alpha_{4,5}$, and the bottom left subplot for each of the along-shore directions (note that omitting the $\cos(\alpha)$ term prevents the collapse of the results).

The linear scaling with $Ri^{1/2}$ confirms that the dimensional $Q_{shore}$ physically depends on both velocity scales via $[nM_g + mW_*]$ as postulated above. For the across-shore directions, this results in (same scaling applies for the along-shore directions but excluding the $\cos(\alpha)$).

$$Ri^{1/2} = a_i \left(\frac{Q_{sc}}{\cos(\alpha_i)}\right) + b_i, \text{ thus} \tag{6}$$

$$Q_{sc} = \left(\frac{Ri^{1/2} - b_i}{a_i}\right) \cos(\alpha_i) = \frac{Q_{shore}}{M_g L_z \|y_n\|}, \text{ and} \tag{7}$$

$$Q_{shore} \propto M_g \cos(\alpha_i) \left(\frac{\frac{W_*}{M_g} - b_i}{a_i}\right) = \cos(\alpha_i) \left[\left(-\frac{b_i}{a_i}\right) M_g + \left(\frac{1}{a_i}\right) W_*\right]. \tag{8}$$

The fitting parameters $a_i$ and $b_i$ determine the coefficients $n_i$ and $m_i$, and the resulting coefficients on the right-hand side of Eq. 8 control the $Q_{shore}$ dependence on $M_g$ and $W_*$, and its behaviour relative to $Ri^{1/2}$. We can then recast these coefficients as $n_i = -\frac{b_i}{a_i}$ and $m_i = \frac{1}{a_i}$. Table 1 summarizes these coefficients $n_i$ and $m_i$ for all $\alpha$. For $\alpha_{1,2}$=0° and 45°, the reported coefficients ($n_{1,2}$=1.003, $m_{1,2}$=4 × 10$^{-3}$) reflect the high dependence on $M_g$ at the expense of $W_*$ when the LSB becomes sea-driven (co-modulated by buoyancy and synoptic pressure simultaneously) or of advected type (modulated exclusively by synoptic pressure). For the $\alpha_{4,5}$=180° and 225°, the reported coefficients ($n_{4,5}$=−1.035, $m_{4,5}$=8 × 10$^{-3}$) imply that when the LSB becomes land-driven along these directions, the dependence on $M_g$ becomes very slightly stronger, but in the opposite direction. The magnitude of $n$ being close to 1 confirms that when $W_* \ll M_g$, the shore flux $\propto \cos(\alpha) M_g$. The magnitude of $m$ then explains the impact of buoyancy in modulating the net flux. For $\alpha_{1,2}$=0° and 45°, $m$ has the same sign as $n$ and thus aids the flux (as also illustrated by the normalized $Q_{sc}$>1), while for $\alpha_{4,5}$=180° and 225°, $m$ is larger and of opposite sign to $n$, revealing a stronger role in damping shore flux (as also illustrated by the normalized $Q_{sc}$<1) due to the persistent very shallow land-driven LSB driving flow at the surface from sea to land.

For the along-shore directions, $\alpha_{3,6}$=90° and 270°, the reported coefficients are drastically different with ($n_3$=−0.048, $m_3$=8 × 10$^{-3}$) and ($n_6$=0.033, $m_6$=4 × 10$^{-3}$) respectively. The resultant LSBs along these directions are all of the canonical type, and therefore $Q_{shore}$ is expected to depend weakly on $M_g$ (hence the small magnitude of $n$ relative to the other angles). On the other hand, $m$ retains the same order of magnitude across angles, implying a more resilient impact of $W_*$. Recall that the directionality of the $X$ Ekman component ($\alpha_3$=90°, land to sea and $\alpha_6$=270°, sea to land) is responsible for marginally enhancing $Q_{shore}^{LS}$ when $\alpha_3$=90° and $Q_{shore}^{SL}$ when $\alpha_6$=270°. Therefore, $n_3$ is negative and $n_6$ is positive (opposing trends). In addition, in the limit of $W_*$= 0 ($M_g$ parallel to shore), Coriolis turning causes a flux rotation averaged across the ABL of about 3 degrees.

*5.4 Advective heat flux at the shore*

We can then define the net advective heat flux $q_{shore}$ (like $Q_{shore}$, but per unit area for interpretation convenience) at the shore interface $X$=40 km ($q_{shore} = q_{shore}^{SL} - q_{shore}^{LS}$) following:

$$q_{shore} = q_{shore}^{SL} - q_{shore}^{LS} = \frac{1}{L_y L_z} \int_0^{L_y} \int_{z_0}^{L_z} U_{shore} \theta_{shore} dz\, dy \qquad (9)$$

A modelled $q_{shore}$ ($q_{shore\text{-}modeled}$, following Eq. 8) can be formulated using the empirical expressions for $Q_{shore-modeled}$ and the average temperature $\theta_{avg}= (\theta_L+\theta_S)/2$ following

$$q_{shore-modeled} = \frac{Q_{shore-modeled}}{L_z \|y_n\|} \theta_{avg} \qquad (10)$$

Fig. 10 shows the agreement between the proposed ($q_{shore\text{-}modeled}$) model (based on the $Q_{shore-modeled}$ model) and the $q_{shore}$ LES results.

The successful scaling here suggests the potential for similar type of models of mass and heat flux to be generalizable for any $\alpha$ (needing only knowledge of the directionality of the synoptic pressure forcing: if it is blowing from land to sea or sea to land), but this needs further testing for variable $W_*$ as well. The net volumetric flux model proposed in Eq. 8 ($Q_{shore-modeled}$) and net heat flux in Eq. 10 could be applied for sea source winds (confidently in the range $\alpha_{sea} \in [-45°, 45°]$) and land source winds (confidently in the range $\alpha_{land} \in [135°, 225°]$). Outside these quadrants, especially in the vicinity of $\alpha_3$=90° and $\alpha_6$=270° where $|V_g| \gg |U_g|$, additional numerical experiments are needed. These models would be particularly useful for parametrizing the unresolved LSB impacts in coarse climate or weather models.

## 6 LSBs categorization and characterization

### 6.1 Criteria for LSBs categorizations

Visual inspection and previous analysis of the thermal circulations flow structures in sections 4 and 5 reveal four types of LSBs (canonical LSB, sea-driven LSB, land-driven LSB, and advected LSB). Therefore, we consider the vertical profile of the shore-normal velocity $U(z)$ at the land-sea interface, $U(z)_{shore}$, with the aid of *k*-means clustering algorithm (Lloyd, 1982) to objectively (without any human visual assessment) delineate these circulations (left subplot of Fig. 11). This autonomous algorithm is provided with $U(z)_{shore}$ as an input, and the number of the desired clusters (= 4 as depicted). Sea-driven LSBs ($\alpha$=0°, $M_g$ (m/s): 0.4 and 0.8; $\alpha$=45°, $M_g$ (m/s): 0.4, 0.8 and 1.2) are correctly grouped in one cluster, and purely advected LSBs ($\alpha$=0°, $M_g$ (m/s): 1.2, 1.6 and 2; $\alpha$=45°, $M_g$ (m/s): 1.6 and 2) are grouped together in another cluster. Land-driven LSBs ($\alpha$=180° and 225°, $M_g$ (m/s): 0.8, 1.2, 1.6 and 2) are similarly grouped in another new cluster. Canonical LSBs are then also recognized for the case with $M_g$=0, as well as all the cases with $\alpha$=270° and 90° for any $M_g$, and for $\alpha$=180° and 225° at $M_g$=0.4.

In the right subplot of Fig. 11, the vertical profile of the advected heat flux at the shore, $U\theta(z)_{shore}$, is considered as an input in the *k*-means clustering (refer also to SI section 4, Fig. S14). The grouping of the clusters exactly matches the results shown in the left subplot.

### 6.2 LSB energetics

Fig. 12 depicts the time and *X-Y-Z* averaged (denoted here by overbar) enstrophy over the analysis domain [10 km-70 km]. It reflects the overall strength of the mean LSB. All results here are normalized with the value of the null synoptic forcing scenario (initial condition). Therefore, the first column at all angles has a value of 1 due to self-normalization. The row corresponding to $\alpha=0°$ shows that the sea-driven LSBs have a high sensitivity to $Ri^{-1}$, with increases in enstrophy by a factor between 2.5 and 3.5. Similarly, the sea-driven LSBs along $\alpha=45°$ show amplification of enstrophy by a factor between 1.8 and 3.5 relative to $M_g=0$. When the sea-driven LSBs shift to advected LSBs type, the normalized enstrophy values drop considerably as the circulations decay, especially for high $M_g$. The row corresponding to $\alpha=180°$ shows that the first perturbation to the $M_g=0$ ($M_g=0.4$ m/s) decreases the enstrophy slightly (similarly for the case $\alpha=225°$: $M_g=0.4$ m/s). These two cases (share features of both land-driven and canonical LSBs types (as depicted in subsection 4.2). As $M_g$ increases for land-driven LSBs, the normalized enstrophy increases to around 1.5. Thus, land-driven LSBs show lower sensitivity to synoptic forcing compared with sea-driven LSBs types. Further increases in $M_g$ along $\alpha=180°$ or $\alpha=225°$ continue to result in land-driven LSB, but also share features of the advected LSBs type especially on the land patch. The latter effect decreases the enstrophy back to values comparable to the enstrophy of the canonical LSB and sometimes less.

The rows corresponding to ($\alpha=270°$ and $90°$), canonical LSBs, indicate that enstrophy fluctuates randomly between 0.5 and 1.2 as $M_g$ is increased. It is worth noting that at the higher $M_g$ value of 2 m/s, as synoptic advection dominates, all cases for any $\alpha$ seem to converge towards a common value of enstrophy, the limit of which would be reached as $Ri$ tends to 0 (neutral wall bounded flow limit). Analyzing the normalized vorticity strength and turbulent kinetic energy (not shown here) revealed the same trends for LSB strength as with the normalized enstrophy.

To summarize, with mild synoptic forcing sea-driven LSBs are characterized by larger secondary circulation energetic levels than the canonical LSBs, while land-driven LSBs have lower energetic levels. As the synoptic pressure gradients strengthens and advection becomes more pronounced, these energetic metrics drop to values much lower than the canonical LSBs.

# 7 Conclusion and implications

A suite of LES cases, spanning thirty combinations of {$M_g$ (m/s): 0.4, 0.8, 1.2, 1.6, 2 and $\alpha$: 0°, 45°, 90°, 180°, 225°, 270°}, reveal the influence of synoptic-scale variability on LSB circulation and flow and transport in coastal zones, with implications for offshore wind energy, coastal urban environments, and city ventilation, among others. The strength of the synoptic forcing $M_g$ and its directionality $\alpha$ jointly modulate a complex interaction between the regional mean flow and the secondary circulations. Along-shore cases, with $V_g=M_g$ ($\alpha=90°$ or $270°$), result in deep thermal circulations that are quite like the canonical LSBs with zero background wind, irrespective of the strength of $M_g$. Across-shore simulations exhibit a shallower circulation cell (sea-driven LSB, when $α=0°$ or $45°$) at low to moderate $M_g$ strength, transitioning to an advected regime with little to no LSB at higher $M_g$. A shallow but persistent circulation (land-driven LSB) develops when $α=180°$ or $225°$. The effect of the $M_g$ strength becomes quite prominent when it is perfectly perpendicular to the shore ($U_g=M_g$) for both scenarios $α=0°$ or $180°$. The oblique cases with weakened $U_g$ ($U_g=V_g=±M_g/2^{1/2}$), when $α=45°$ or $225°$, result in similar patterns to the ones at $α=0°$ or $180°$ with a comparable $U_g$, hinting at the direct significance of $U_g$ in driving the circulation. The open debate in the literature on the synoptic wind speed at which LSBs disappear (see introduction) may thus be caused by the important, but maybe neglected, role of the wind angle $α$.

There is asymmetry in the land-to-sea versus sea-to-land synoptic forcing: sea-driven LSBs render the sea neutral and the land highly unstable, while land-driven LSBs drive the land neutral and the sea highly stable. As a result, sea-driven LSBs affect the $SB_{depth}$ non-monotonically as $M_g$ is increased (first decreasing it, then increasing it), whereas $SB_{depth}$ decays in a sharp monotonic sense until it plateaus with land-driven LSBs. A scaling analysis of the net volumetric flux postulating $Q_{shore} \sim nM_g + mW_*$ is proposed and evaluated, and reveals a dependence of the normalized flux $Q_{sc}$ on $Ri^{½}$ The determined $n$ and $m$ coefficients (from LES) reflect the physical implications of the identified LSBs, and support the generalizability of these relations for different $α$'s (though additional work is needed). Similarly, the net advective heat flux at the shore, $q_{shore}$, can be modelled using the derived empirical relations for $Q_{shore}$, and the results show good agreement with LES-derived heat flux.

The vertical profiles of either the shore-normal velocity $U(z)$ or the advected heat flux $Uθ(z)$ at the shore are then used, with the aid of $k$-means, to delineate these circulations based on a regime space diagram with respect to $Ri^{-1}$ and $α$. The various regimes (canonical LSB, sea-driven LSB, land-driven LSB, and advected LSB) are shown to define distinctive ranges of enstrophy, but they might overlap when features of two LSB types coexist.

The shore flux models and scaling developed here would be beneficial as a first step in developing parameterization schemes for general climate models (GCMs) or coarse weather prediction models. Capturing the effect of these LSBs, which only span <50 km at most (unresolved in coarse models), is essential since they have major impacts on rainfall, temperature, wind, and humidity near coasts. Coastal winds also control upwelling in the ocean, which then alters sea surface temperature and feeds back to alter land-sea breezes, in addition to its ecological effects. The implications on offshore wind power of these dynamics are emerging as another major motivation to better understand the timing and scales of these unsteady flows (Howland *et al.*, 2020). Thus, a follow up to this study is now to impose time-varying surface temperature boundary condition over land. Diurnal analysis would offer new insights on the competition between the constant pressure forcing, buoyancy from the transient temperature difference between the two patches, and the inertial/memory effects of the LSBs. The inertia of these circulations is large and will play a key role in transient dynamics, something the present paper did not consider.

**Acknowledgments**

M.A. and E.B.Z. are supported by the Cooperative Institute for Modeling the Earth System at Princeton University under Award NA18OAR4320123 from the National Oceanic and Atmospheric Administration, and by the US National Science Foundation under award number AGS 2128345. The LES simulations are conducted on the Cheyenne supercomputer of the National Center for Atmospheric Research (doi:10.5065/D6RX99HX) under projects UPRI0007 and UPRI0021.

**Data availability**

Data will be made available upon request.

**Conflict of interest disclosure**

The authors declare no conflict of interest.

**Supporting Information**

*Supporting Section 1 (Sensitivity analysis to the numerical grid resolution: resolution and Y sensitivity, averaging time influence on the analysed statistics)*

Fig. S1 Pseudocolour plots of along-shore, time averaged stream-wise velocity $\langle U \rangle_{y,t}$ through an *X-Z* slice of the analysis domain for the case of $M_g$=0 m/s (control, top), (resolution sensitivity, middle), (*Y* sensitivity, bottom)

Fig. S2 Pseudocolour plots of along-shore, time averaged temperature $\langle \theta \rangle_{y,t}$ through an *X-Z* slice of the analysis domain for the case of $M_g$=0 m/s (control, top), (resolution sensitivity, middle), (*Y* sensitivity, bottom)

Fig. S3 Pseudocolour plots of along-shore, time averaged stream-wise velocity $\langle U \rangle_{y,t}$ through an *X-Z* slice of the analysis domain for the case of $\alpha$=180° and $M_g$=1.2 m/s (control, top), (resolution sensitivity, middle), (*Y* sensitivity, bottom)

Fig. S4 Pseudocolour plots of along-shore, time averaged temperature $\langle \theta \rangle_{y,t}$ through an *X-Z* slice of the analysis domain for the case of $\alpha$=180° and $M_g$=1.2 m/s (control, top), (resolution sensitivity, middle), (*Y* sensitivity, bottom).

Table S1 Comparing the analysed LES quantities of the resolution and *Y* sensitivity simulations relative to the control simulation ($\alpha$=180° and $M_g$=1.2 m/s setup)

Table S2 Comparing the analysed LES quantities of the resolution and *Y* sensitivity simulations relative to the control simulation ($M_g$=0 m/s setup)

Table S3 Comparing the analysed LES quantities of the *Y* sensitivity simulations over different periods ($\alpha$=180° and $M_g$=1.2 m/s setup)

Table S4 Comparing the analysed LES quantities of the *Y* sensitivity simulations relative to the control simulation ($\alpha$=180° and $M_g$=1.2 m/s setup) over different periods

Table S5 Comparing the analysed LES quantities of the resolution sensitivity simulations relative to the control simulation ($\alpha$=180° and $M_g$=1.2 m/s setup) over different periods

*Supporting Section 2 (Thermal circulations)*

Figs. S5, S7, S9, S11: Pseudocolour plots of along-shore, time averaged stream-wise velocity $\langle U \rangle_{y,t}$ through an *X-Z* slice of the analysis domain.

Figs. S6, S8, S10, S12: Pseudocolour plots of along-shore, time averaged temperature $\langle \theta \rangle_{y,t}$ through an *X-Z* slice of the analysis domain.

*Supporting Section 3 (Surface friction velocity)*

Fig. S13 A scatter plot of the surface friction velocities over the two patches for all $\alpha$'s with increasing $Ri^{-1}$ (sea: left, land: right)

*Supporting Section 4 (Stream-wise advective heat flux profile)*

Fig. S14 Shore stream-wise advective heat flux profiles for all $\alpha$'s with increasing $M_g$

# References

Antonelli, M. and Rotunno, R. (2007) 'Large-eddy simulation of the onset of the sea breeze', *Journal of the Atmospheric Sciences* [Preprint]. Available at: https://doi.org/10.1175/2007JAS2261.1.

Arritt, R.W. (1993a) 'Effects of the large-scale flow on characteristic features of the sea breeze', *Journal of Applied Meteorology* [Preprint]. Available at: https://doi.org/10.1175/1520-0450(1993)032<0116:EOTLSF>2.0.CO;2.

Arritt, R.W. (1993b) 'Effects of the large-scale flow on characteristic features of the sea breeze', *Journal of Applied Meteorology* [Preprint]. Available at: https://doi.org/10.1175/1520-0450(1993)032<0116:EOTLSF>2.0.CO;2.

Atkins, N.T. and Wakimoto, R.M. (1997) 'Influence of the synoptic-scale flow on sea breezes observed during CaPE', *Monthly Weather Review* [Preprint]. Available at: https://doi.org/10.1175/1520-0493(1997)125<2112:IOTSSF>2.0.CO;2.

Atkison, B.W. (1995) 'Sea breeze and local wind. By J. E. Simpson. Cambridge University Press. 1994. Pp. 234. Price £29.95 (hardback). ISBN 0 521 452112', *Quarterly Journal of the Royal Meteorological Society* [Preprint]. Available at: https://doi.org/10.1002/qj.49712152315.

Azorin-Molina, C. and Chen, D. (2009) 'A climatological study of the influence of synoptic-scale flows on sea breeze evolution in the Bay of Alicante (Spain)', *Theoretical and Applied Climatology* [Preprint]. Available at: https://doi.org/10.1007/s00704-008-0028-2.

Banta, R.M., Olivier, L.D. and Levinson, D.H. (1993) 'Evolution of the Monterey Bay Sea-Breeze Layer As Observed by Pulsed Doppler Lidar', *Journal of the Atmospheric Sciences* [Preprint]. Available at: https://doi.org/10.1175/1520-0469(1993)050<3959:eotmbs>2.0.co;2.

Bechtold, P., Pinty, J.P. and Mascart, P. (1991) 'A numerical investigation of the influence of large-scale winds on sea-breeze- and inland-breeze-type circulations', *Journal of Applied Meteorology* [Preprint]. Available at: https://doi.org/10.1175/1520-0450(1991)030<1268:ANIOTI>2.0.CO;2.

Bou-Zeid, E. *et al.* (2020) 'The Persistent Challenge of Surface Heterogeneity in Boundary-Layer Meteorology: A Review', *Boundary-Layer Meteorology* [Preprint]. Available at: https://doi.org/10.1007/s10546-020-00551-8.

Bou-Zeid, E., Meneveau, C. and Parlange, M. (2005) 'A scale-dependent Lagrangian dynamic model for large eddy simulation of complex turbulent flows', *Physics of Fluids* [Preprint]. Available at: https://doi.org/10.1063/1.1839152.

Cana, L., Grisolía-Santos, D. and Hernández-Guerra, A. (2020) 'A Numerical Study of a Sea Breeze at Fuerteventura Island, Canary Islands, Spain', *Boundary-Layer Meteorology* [Preprint]. Available at: https://doi.org/10.1007/s10546-020-00506-z.

Chen, G. *et al.* (2015) 'Toward improved forecasts of sea-breeze horizontal convective rolls at super high resolutions. Part II: The impacts of land use and buildings', *Monthly Weather Review* [Preprint]. Available at: https://doi.org/10.1175/MWR-D-14-00230.1.

Cioni, G. and Hohenegger, C. (2018) 'A simplified model of precipitation enhancement over a heterogeneous surface', *Hydrology and Earth System Sciences* [Preprint]. Available at: https://doi.org/10.5194/hess-22-3197-2018.

Collier, C.G. (2006) 'The impact of urban areas on weather', *Quarterly Journal of the Royal Meteorological Society* [Preprint]. Available at: https://doi.org/10.1256/qj.05.199.

Corner, N.T. and McKendry, I.G. (1993) 'Observations and numerical modelling of lake Ontario breezes', *Atmosphere - Ocean* [Preprint]. Available at: https://doi.org/10.1080/07055900.1993.9649482.

Crosman, E.T. and Horel, J.D. (2010a) 'Sea and Lake Breezes: A Review of Numerical Studies', *Boundary-Layer Meteorology* [Preprint]. Available at: https://doi.org/10.1007/s10546-010-9517-9.

Crosman, E.T. and Horel, J.D. (2010b) 'Sea and Lake Breezes: A Review of Numerical Studies', *Boundary-Layer Meteorology* [Preprint]. Available at: https://doi.org/10.1007/s10546-010-9517-9.

D. Hinrichsen (1997) 'Coastal Waters of the World: Trends, Threats, and Strategies', *Colonial Waterbirds*, 20(3), p. 630. Available at: https://doi.org/10.2307/1521623.

Dalu, G.A. and Pielke, R.A. (1989) 'An analytical study of the sea breeze', *Journal of the Atmospheric Sciences* [Preprint]. Available at: https://doi.org/10.1175/1520-0469(1989)046<1815:AASOTS>2.0.CO;2.

Delage, Y. and Taylor, P.A. (1970) 'Numerical studies of heat island circulations', *Boundary-Layer Meteorology* [Preprint]. Available at: https://doi.org/10.1007/BF00185740.

Drobinski, P. *et al.* (2006) 'Variability of three-dimensional sea breeze structure in southern France: Observations and evaluation of empirical scaling laws', *Annales Geophysicae* [Preprint]. Available at: https://doi.org/10.5194/angeo-24-1783-2006.

Estoque, M.A. (1962) 'The Sea Breeze as a Function of the Prevailing Synoptic Situation', *Journal of the Atmospheric Sciences* [Preprint]. Available at: https://doi.org/10.1175/1520-0469(1962)019<0244:tsbaaf>2.0.co;2.

Finkele, K. *et al.* (1995) 'A complete sea-breeze circulation cell derived from aircraft observations', *Boundary-Layer Meteorology* [Preprint]. Available at: https://doi.org/10.1007/BF00711261.

Finnigan, J.J., Shaw, R.H. and Patton, E.G. (2009) 'Turbulence structure above a vegetation canopy', *Journal of Fluid Mechanics* [Preprint]. Available at: https://doi.org/10.1017/S0022112009990589.

Garratt, J.R. (1990) 'The internal boundary layer - A review', *Boundary-Layer Meteorology* [Preprint]. Available at: https://doi.org/10.1007/BF00120524.

Ghannam, K. and Bou-Zeid, E. (2021) 'Baroclinicity and directional shear explain departures from the logarithmic wind profile', *Quarterly Journal of the Royal Meteorological Society* [Preprint]. Available at: https://doi.org/10.1002/qj.3927.

Gilliam, R.C., Raman, S. and Niyogi, D.D.S. (2004) 'Observational and numerical study on the influence of large-scale flow direction and coastline shape on sea-breeze evolution', *Boundary-Layer Meteorology* [Preprint]. Available at: https://doi.org/10.1023/B:BOUN.0000016494.99539.5a.

Grisogono, B., Ström, L. and Tjernström, M. (1998) 'Small-scale variability in the coastal atmospheric boundary layer', *Boundary-Layer Meteorology* [Preprint]. Available at: https://doi.org/10.1023/A:1000933822432.

Hadfield, M.G., Cotton, W.R. and Pielke, R.A. (1991) 'Large-eddy simulations of thermally forced circulations in the convective boundary layer. Part I: A small-scale circulation with zero wind', *Boundary-Layer Meteorology* [Preprint]. Available at: https://doi.org/10.1007/BF00119714.

Hadfield, M.G., Cotton, W.R. and Pielke, R.A. (1992) 'Large-eddy simulations of thermally forced circulations in the convective boundary layer. Part II: The effect of changes in wavelength and wind speed', *Boundary-Layer Meteorology* [Preprint]. Available at: https://doi.org/10.1007/BF00120235.

Haurwitz (1947) 'COMMENTS ON THE SEA-BREEZE CIRCULATION', *Journal of the Atmospheric Sciences* [Preprint]. Available at: https://doi.org/https://doi.org/10.1175/1520-0469(1947)004<0001:COTSBC>2.0.CO;2.

Howland, M.F. *et al.* (2020) 'Influence of atmospheric conditions on the power production of utility-scale wind turbines in yaw misalignment', *Journal of Renewable and Sustainable Energy* [Preprint]. Available at: https://doi.org/10.1063/5.0023746.

Huang, J. and Bou-Zeid, E. (2013) 'Turbulence and Vertical Fluxes in the Stable Atmospheric Boundary Layer. Part I: A Large-Eddy Simulation Study', *Journal of the Atmospheric Sciences* [Preprint]. Available at: https://doi.org/10.1175/jas-d-12-0167.1.

Jiang, P. *et al.* (2017) 'Interaction between turbulent flow and sea breeze front over urban-like coast in large-eddy simulation', *Journal of Geophysical Research* [Preprint]. Available at: https://doi.org/10.1002/2016JD026247.

Kumar, V. *et al.* (2006) 'Large-eddy simulation of a diurnal cycle of the atmospheric boundary layer: Atmospheric stability and scaling issues', *Water Resources Research* [Preprint]. Available at: https://doi.org/10.1029/2005WR004651.

LaCasse, K.M. *et al.* (2008) 'The impact of high-resolution sea surface temperatures on the simulated nocturnal Florida marine boundary layer', *Monthly Weather Review* [Preprint]. Available at: https://doi.org/10.1175/2007MWR2167.1.

Li, Y.K. and Chao, J.P. (2016) 'An analytical solution for three-dimensional sea-land breeze', *Journal of the Atmospheric Sciences* [Preprint]. Available at: https://doi.org/10.1175/JAS-D-14-0329.1.

Lloyd, S.P. (1982) 'Least Squares Quantization in PCM', *IEEE Transactions on Information Theory* [Preprint]. Available at: https://doi.org/10.1109/TIT.1982.1056489.

Lund, T.S., Wu, X. and Squires, K.D. (1998) 'Generation of Turbulent Inflow Data for Spatially-Developing Boundary Layer Simulations', *Journal of Computational Physics* [Preprint]. Available at: https://doi.org/10.1006/jcph.1998.5882.

Miller, M.J. and Thorpe, A.J. (1981) 'Radiation conditions for the lateral boundaries of limited-area numerical models', *Quarterly Journal of the Royal Meteorological Society* [Preprint]. Available at: https://doi.org/10.1002/qj.49710745310.

Miller, S.T.K. *et al.* (2003) 'Sea breeze: Structure, forecasting, and impacts', *Reviews of Geophysics* [Preprint]. Available at: https://doi.org/10.1029/2003RG000124.

Munters, W., Meneveau, C. and Meyers, J. (2016) 'Turbulent Inflow Precursor Method with Time-Varying Direction for Large-Eddy Simulations and Applications to Wind Farms', *Boundary-Layer Meteorology* [Preprint]. Available at: https://doi.org/10.1007/s10546-016-0127-z.

Musk, L.F. (1979) 'Book reviews: Dutton, J. A. 1976: The ceaseless wind, an introduction to the theory of atmospheric motion. New York: McGraw-Hill. xv+579 pp. £20.75', *Progress in Physical Geography: Earth and Environment* [Preprint]. Available at: https://doi.org/10.1177/030913337900300410.

Orlanski, I. (1976) 'A simple boundary condition for unbounded hyperbolic flows', *Journal of Computational Physics* [Preprint]. Available at: https://doi.org/10.1016/0021-9991(76)90023-1.

Orszag, S.A. and Pao, Y.H. (1975) 'Numerical computation of turbulent shear flows', *Advances in Geophysics* [Preprint]. Available at: https://doi.org/10.1016/S0065-2687(08)60463-X.

Pope, S.B. (2001) 'Turbulent Flows', *Measurement Science and Technology* [Preprint]. Available at: https://doi.org/10.1088/0957-0233/12/11/705.

Porson, A., Steyn, D.G. and Schayes, G. (2007) 'Sea-breeze scaling from numerical model simulations, Part I: Pure sea breezes', *Boundary-Layer Meteorology* [Preprint]. Available at: https://doi.org/10.1007/s10546-006-9090-4.

Qian, T., Epifanio, C.C. and Zhang, F. (2009) 'Linear theory calculations for the sea breeze in a background wind: The equatorial case', *Journal of the Atmospheric Sciences* [Preprint]. Available at: https://doi.org/10.1175/2008JAS2851.1.

Reible, D.D., Simpson, J.E. and Linden, P.F. (1993) 'The sea breeze and gravity-current frontogenesis', *Quarterly Journal of the Royal Meteorological Society* [Preprint]. Available at: https://doi.org/10.1002/qj.49711950902.

Rotunno, R. (1983) 'On the linear theory of the land and sea breeze.', *Journal of the Atmospheric Sciences* [Preprint]. Available at: https://doi.org/10.1175/1520-0469(1983)040<1999:OTLTOT>2.0.CO;2.

Rotunno, R. *et al.* (1996) 'Meeting summary - Coastal meteorology and oceanography: Report of the third prospectus development team of the U.S. Weather Research Program to NOAA and NSF', in *Bulletin of the American Meteorological Society*. Available at: https://doi.org/10.1175/1520-0477-77.7.1578.

Segal, M. *et al.* (1997) 'Small Lake Daytime Breezes: Some Observational and Conceptual Evaluations', *Bulletin of the American Meteorological Society* [Preprint]. Available at: https://doi.org/10.1175/1520-0477(1997)078<1135:SLDBSO>2.0.CO;2.

Sills, D.M.L. *et al.* (2011) 'Lake breezes in the southern Great Lakes region and their influence during BAQS-Met 2007', *Atmospheric Chemistry and Physics* [Preprint]. Available at: https://doi.org/10.5194/acp-11-7955-2011.

Simpson, J.E. (1969) 'A comparison between laboratory and atmospheric density currents', *Quarterly Journal of the Royal Meteorological Society* [Preprint]. Available at: https://doi.org/10.1002/qj.49709540609.

Spalart, P.R. (1988) 'Direct simulation of a turbulent boundary layer up to R$\theta$= 1410', *Journal of Fluid Mechanics* [Preprint]. Available at: https://doi.org/10.1017/S0022112088000345.

Stevens, R.J.A.M., Graham, J. and Meneveau, C. (2014) 'A concurrent precursor inflow method for Large Eddy Simulations and applications to finite length wind farms', *Renewable Energy* [Preprint]. Available at: https://doi.org/10.1016/j.renene.2014.01.024.

Steyn, D.G. (2003) 'Scaling the vertical structure of sea breezes revisited', *Boundary-Layer Meteorology* [Preprint]. Available at: https://doi.org/10.1023/A:1021568117280.

Sullivan, P.P. *et al.* (2020) 'Marine boundary layers above heterogeneous SST: Across-front winds', *Journal of the Atmospheric Sciences* [Preprint]. Available at: https://doi.org/10.1175/JAS-D-20-0062.1.

Sullivan, P.P. *et al.* (2021) 'Marine boundary layers above heterogeneous SST: Alongfront winds', *Journal of the Atmospheric Sciences* [Preprint]. Available at: https://doi.org/10.1175/JAS-D-21-0072.1.

Tijm, A.B.C., Van Delden, A.J. and Holtslag, A.A.M. (1999) 'The inland penetration of sea breezes', *Contributions to Atmospheric Physics* [Preprint].

Weiming Sha, Kawamura, T. and Ueda, H. (1991) 'A numerical study on sea/land breezes as a gravity current: Kelvin- Helmholtz billows and inland penetration of the sea-breeze front', *Journal of the Atmospheric Sciences* [Preprint]. Available at: https://doi.org/10.1175/1520-0469(1991)048<1649:ansosb>2.0.co;2.

Yang, X. (1991) 'A study of nonhydrostatic effects in idealized sea breeze systems', *Boundary-Layer Meteorology* [Preprint]. Available at: https://doi.org/10.1007/BF00119419.

Yoshikado, H. (1992) 'Numerical study of the daytime urban effect and its interaction with the sea breeze', *Journal of Applied Meteorology* [Preprint]. Available at: https://doi.org/10.1175/1520-0450(1992)031<1146:nsotdu>2.0.co;2.

Zhong, S. and Takle, E.S. (1993) 'The Effects of Large-Scale Winds on the Sea–Land-Breeze Circulations in an Area of Complex Coastal Heating', *Journal of Applied Meteorology*, 32(7), pp. 1181–1195. Available at: https://doi.org/10.1175/1520-0450(1993)032<1181:TEOLSW>2.0.CO;2.

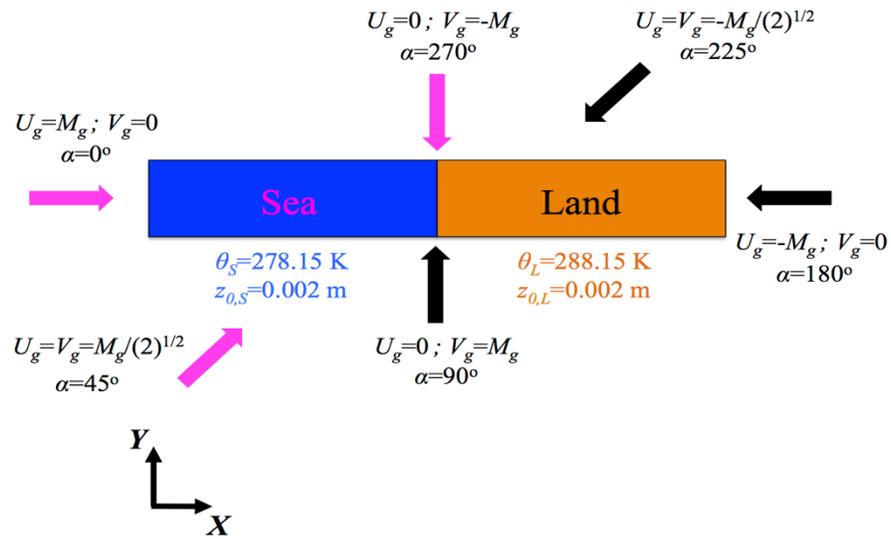

Fig. 1 Schematic diagram of the sea-land configuration. Arrow colours indicate the direction of the surface wind in the absence of a thermal contrast, just due to Ekman turning: the cases where the wind would be blowing from the sea are in magenta, while those blowing from the land are in black

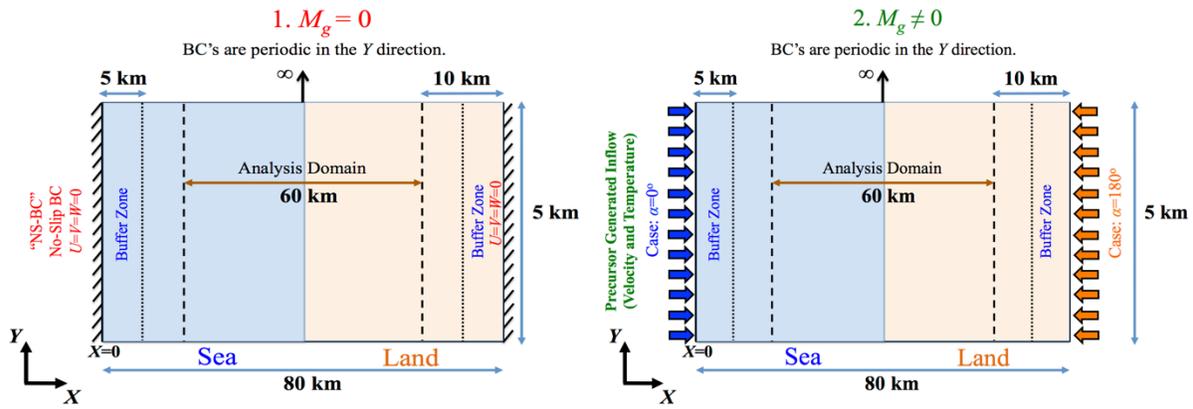

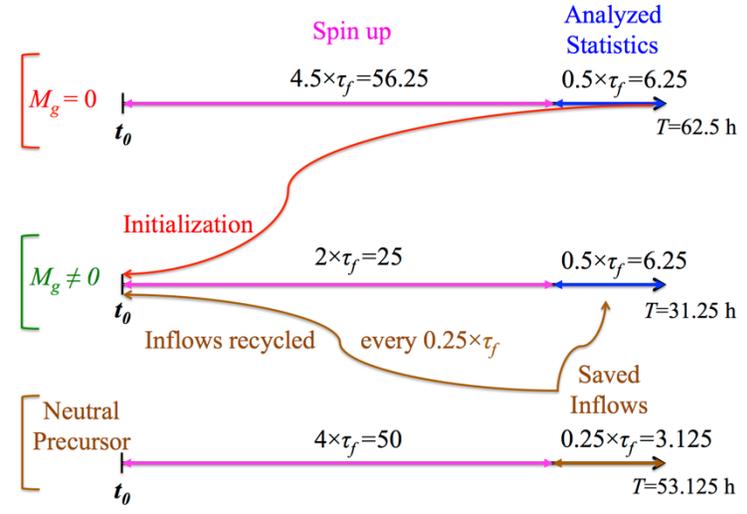

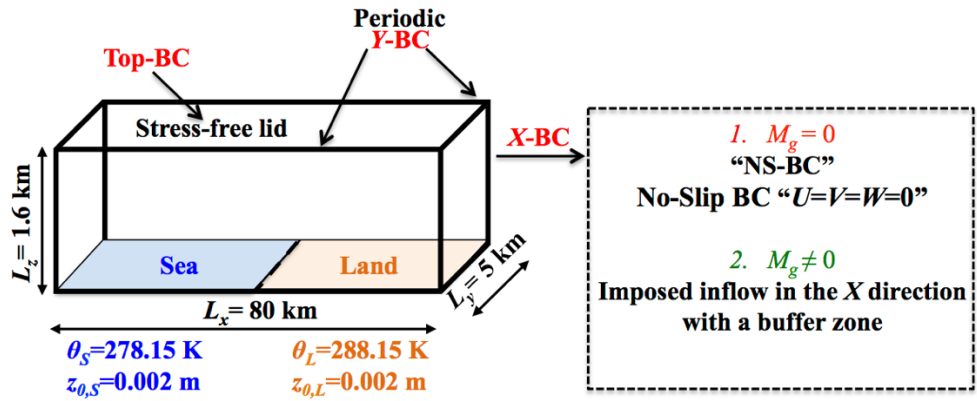

Fig. 2 Top view of the domain for the two scenarios' along-shore boundary conditions ($M_g$=0, top left) and ($M_g\neq0$ top right). Schematic of spin-up, initialization, and analyses periods for all simulations (middle panel). Three dimensional schematic diagram of the sea-land heterogeneous simulation (bottom)

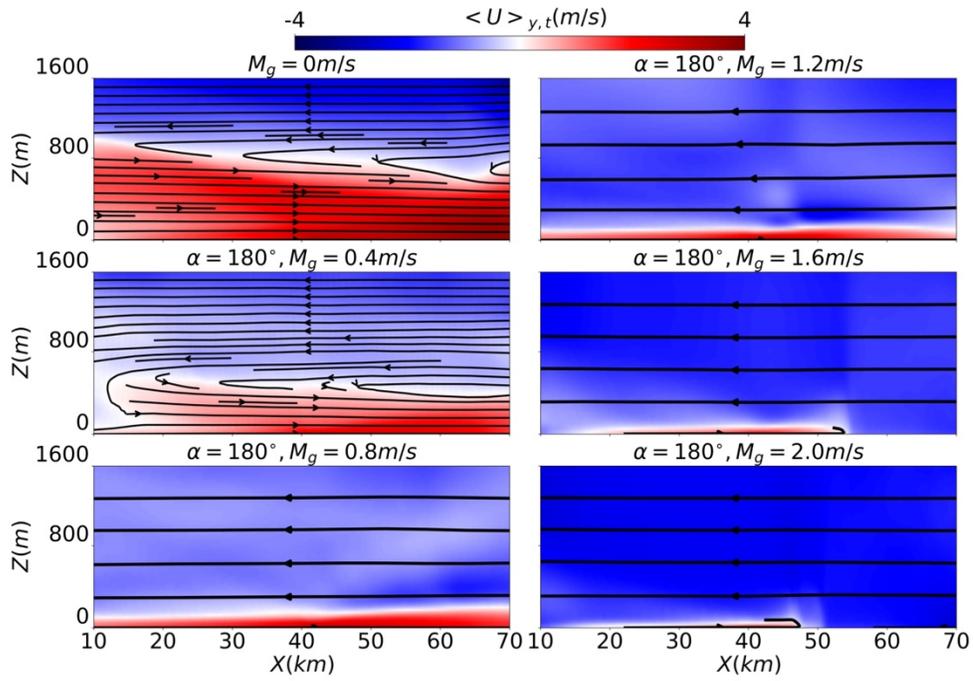

Fig. 3 Pseudocolour plots of along-shore, mean stream-wise velocity $\langle U \rangle_{y,t}$ through an *X-Z* slice of the analysis domain. The series of subplots correspond to $\alpha=180°$ with increasing $M_g$

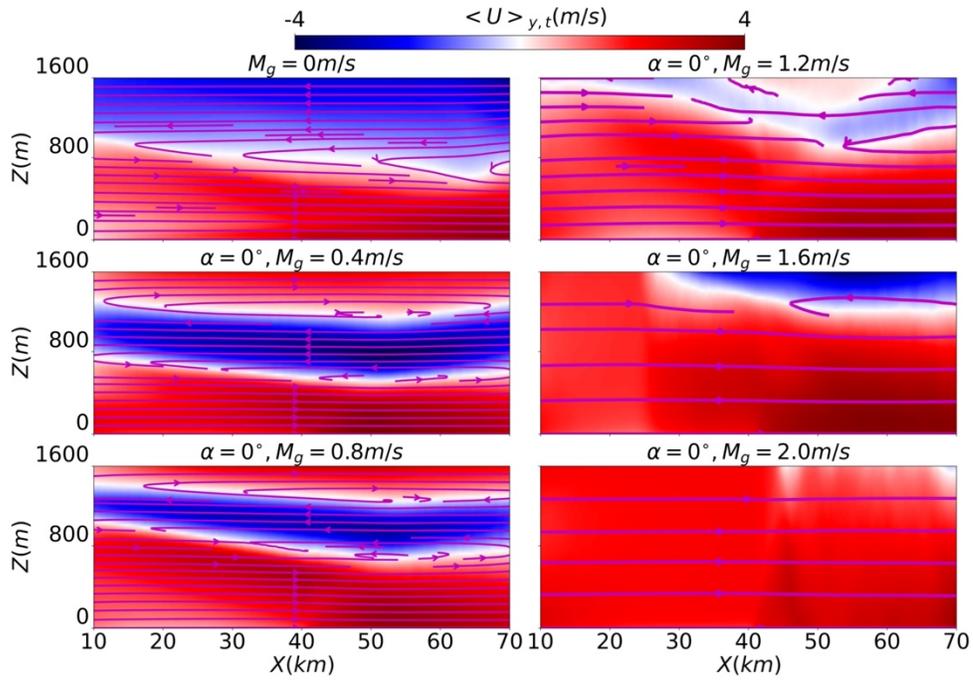

Fig. 4 Pseudocolour plots of along-shore mean stream-wise velocity $\langle U \rangle_{y,t}$ through an *X-Z* slice of the analysis domain. The series of subplots correspond to $\alpha=0°$ with increasing $M_g$

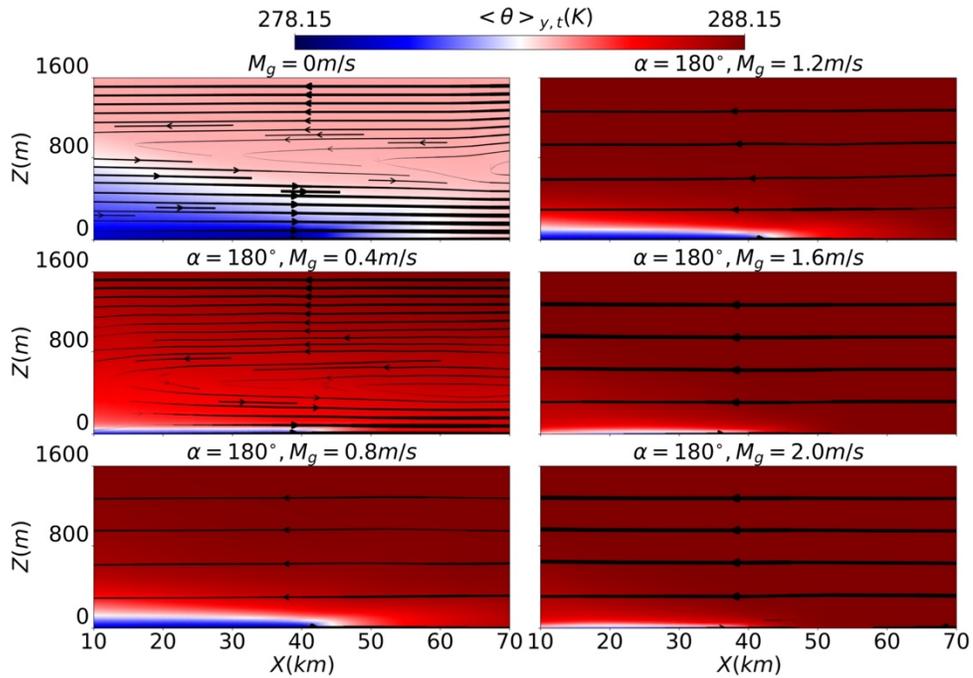

Fig. 5 Pseudocolour plots of along-shore, mean temperature $\langle\theta\rangle_{y,t}$ through an *X-Z* slice of the analysis domain. The series of subplots correspond to $\alpha=180°$ with increasing $M_g$. The streamlines are drawn with varied linewidth to reflect the wind speed in each subplot

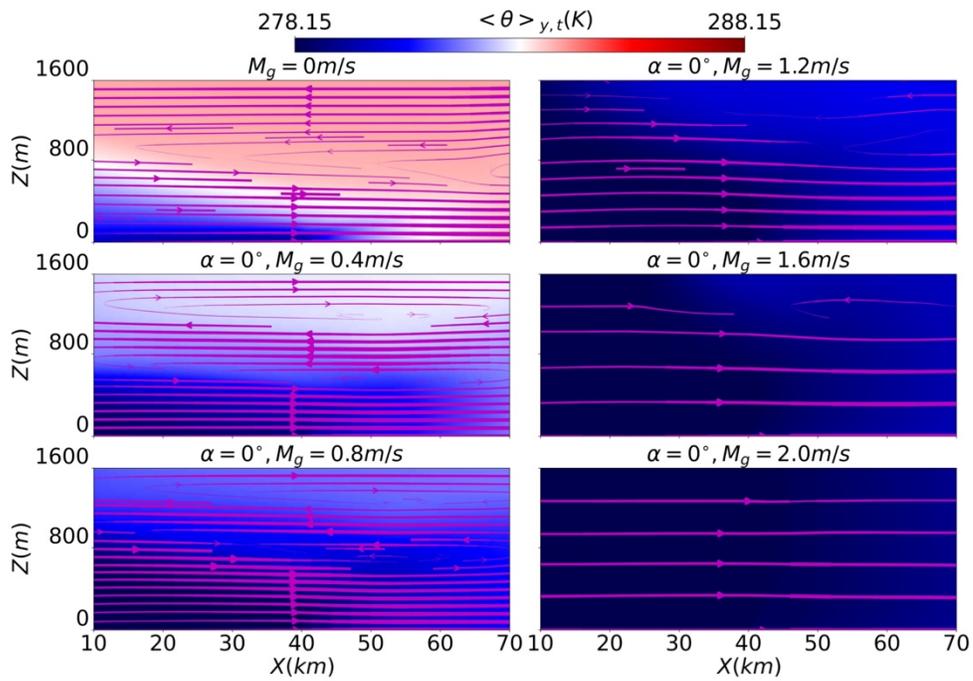

Fig. 6 Pseudocolour plots of along-shore, mean temperature $\langle\theta\rangle_{y,t}$ through an *X-Z* slice of the analysis domain. The series of subplots correspond to $\alpha=0°$ with increasing $M_g$. The streamlines are drawn with varied linewidth to reflect the wind speed in each subplot

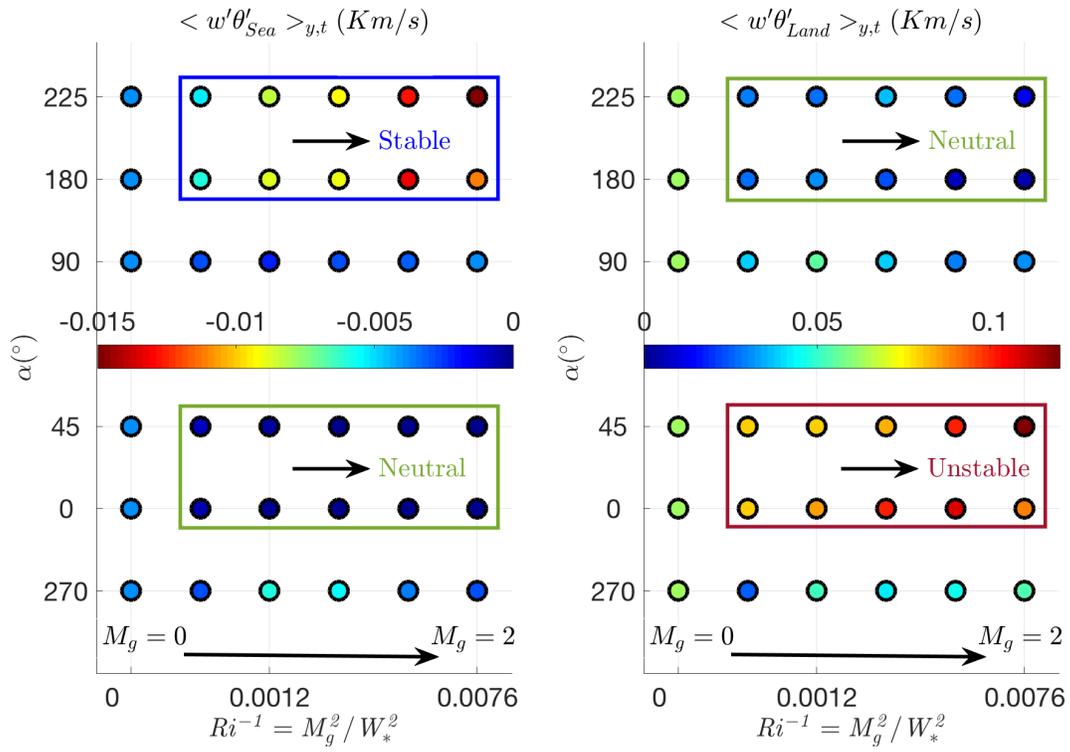

Fig. 7 The surface heat fluxes (K m/s) over the two patches for all $\alpha$'s with increasing $R_i^{-1}$ (sea: left, land: right; flow from land to sea in top panel and from sea to land in bottom panel)

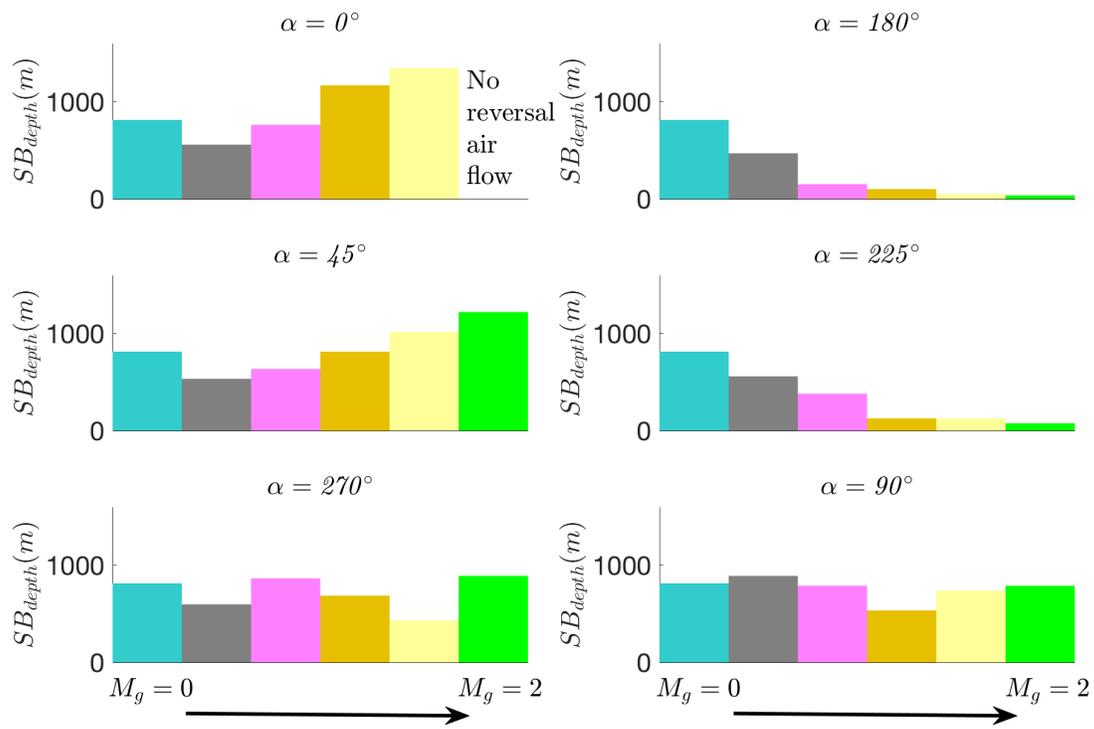

Fig. 8 The sea breeze depth at the shore for all $\alpha$'s with increasing $M_g$

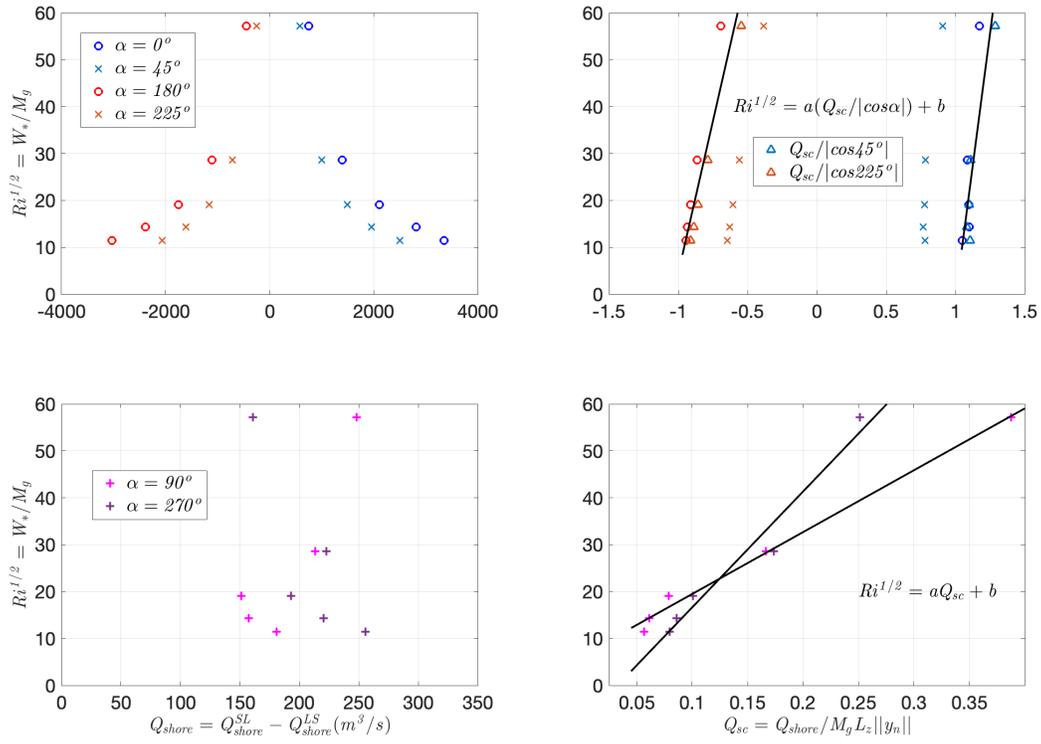

Fig. 9 Shore volumetric flux ($Q_{shore}$) variation with respect to $Ri^{1/2}$ for every $\alpha$ (left panel), and a normalized version $Q_{sc}$ (right panel) with the associated linear models (black curves)

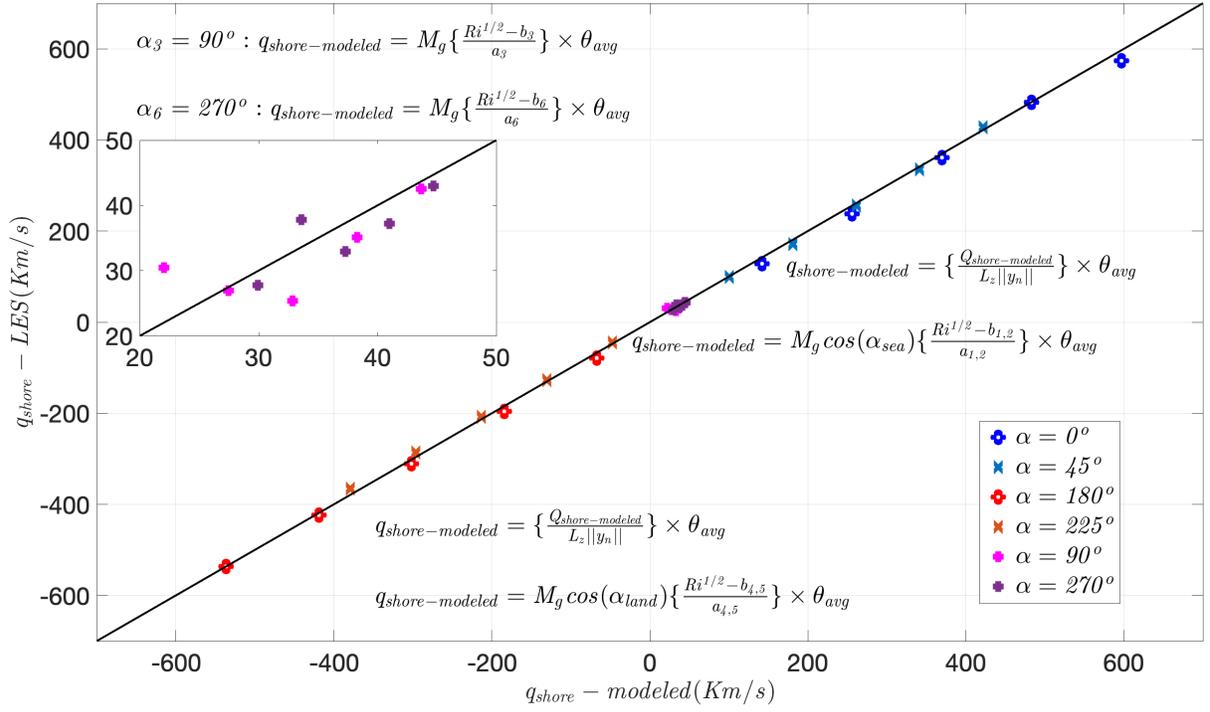

Fig. 10 The modelled $q_{shore}$ versus the LES $q_{shore}$ for all $\alpha$'s with increasing $M_g$.
One-to-one line is shown as reference (solid black line)

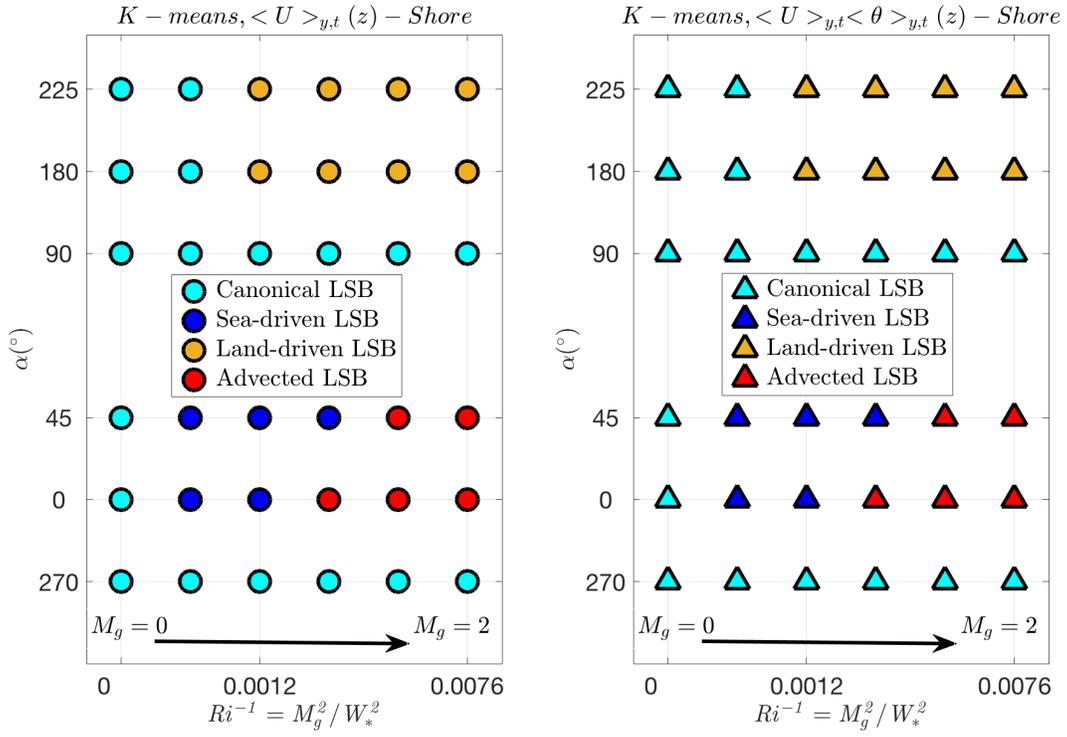

Fig. 11 Classification of all the simulation cases using the *k*-means algorithm with ($U(z)_{shore}$, left) and ($U\theta(z)_{shore}$, right)

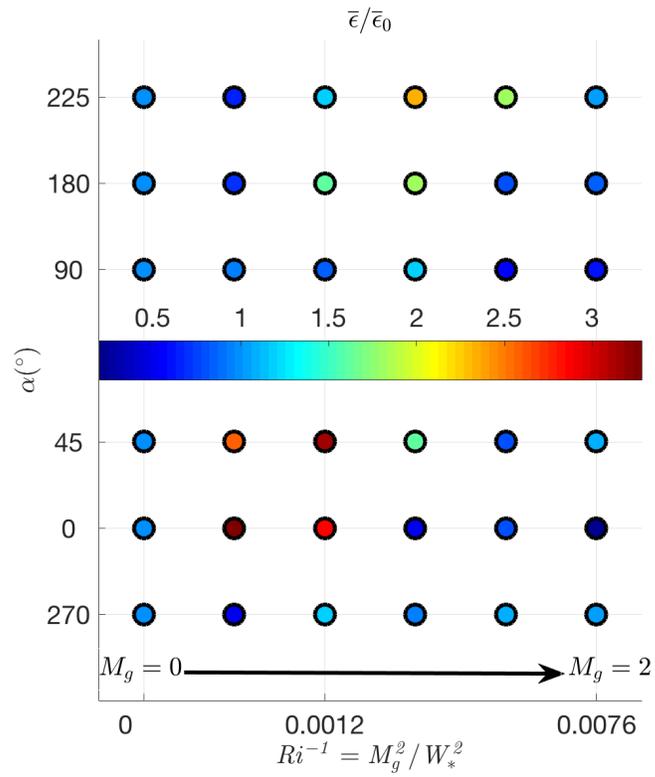

Fig. 12 Time and *X-Y-Z* averaged enstrophy for all *α*'s with increasing $Ri^{-1}$

Table 1 The fitting parameters *a, b, n* and *m* for the flow models for all wind angles

| Fitting parameters | Across-Shore (Sea) $\alpha_{1,2} = 0°, 45°$ $Ri^{1/2} = a_{1,2}\left(\frac{Q_{sc}}{\cos(\alpha_{1,2})}\right) + b_{1,2}$ | Across-Shore (Land) $\alpha_{4,5} = 180°, 225°$ $Ri^{1/2} = a_{4,5}\left(\frac{Q_{sc}}{\cos(\alpha_{4,5})}\right) + b_{4,5}$ | Along-Shore $\alpha_{3,6} = 90°, 270°$ $Ri^{1/2} = a_3 Q_{sc} + b_3$ $Ri^{1/2} = a_6 Q_{sc} + b_6$ |
|---|---|---|---|
| $a_i$ | $a_{1,2} = 225.8$ | $a_{4,5} = 129.6$ | $a_3 = 131.9$ $a_6 = 247$ |
| $b_i$ | $b_{1,2} = -226.5$ | $b_{4,5} = 134.1$ | $b_3 = 6.3$ $b_6 = -8.1$ |
| $\binom{n}{m}_i$ | $\binom{n}{m}_{1,2} = \binom{1.003}{4 \times 10^{-3}}$ | $\binom{n}{m}_{4,5} = \binom{-1.035}{8 \times 10^{-3}}$ | $\binom{n}{m}_3 = \binom{-0.048}{8 \times 10^{-3}}$ $\binom{n}{m}_6 = \binom{0.033}{4 \times 10^{-3}}$ |
| $R_i^2$ $(R^2)$ | $R_{1,2}^2 = 0.732$ | $R_{4,5}^2 = 0.883$ | $R_3^2 = 0.992$ $R_6^2 = 0.959$ |

# Supporting Information:
# The Influence of Synoptic Wind on Coastal Circulation Dynamics

Mohammad Allouche[a], Elie Bou-Zeid[a,*], Juho Iipponen[b]

[a] *Department of Civil and Environmental Engineering, Princeton University, New Jersey*

[b] *Program in Atmospheric and Oceanic Sciences, Princeton University, New Jersey*

*Corresponding author: Elie Bou-Zeid ebouzeid@princeton.edu*

## 8 Sensitivity analysis to the numerical grid resolution: resolution and *Y* sensitivity, averaging time influence on the analysed statistics

Figures S-1 and S-2 show, respectively, pseudocolour plots of along-shore, time averaged stream-wise velocity $\langle U \rangle_{y,t}$ and temperature $\langle \theta \rangle_{y,t}$ through an *X-Z* slice of the analysis domain (control, top), (resolution sensitivity, middle), (*Y* sensitivity, bottom) for the $M_g = 0$ m/s scenario. The large-scale features of the LSBs of the sensitivity simulations are similar to like the control case.

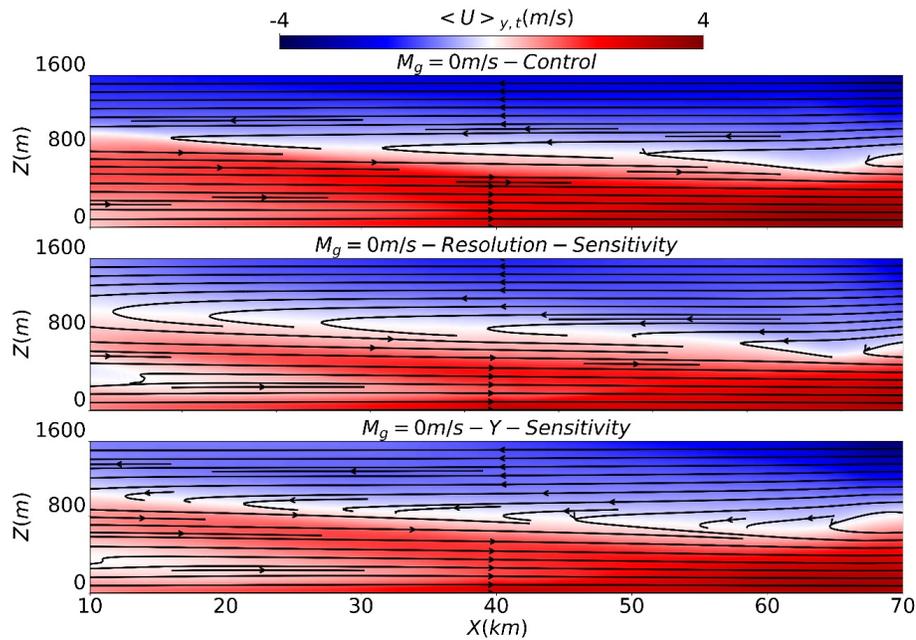

Figure S1 Pseudocolour plots of along-shore, time averaged stream-wise velocity $\langle U \rangle_{y,t}$ through an *X-Z* slice of the analysis domain for the case of $M_g=0$ m/s (control, top), (resolution sensitivity, middle), (*Y* sensitivity, bottom)

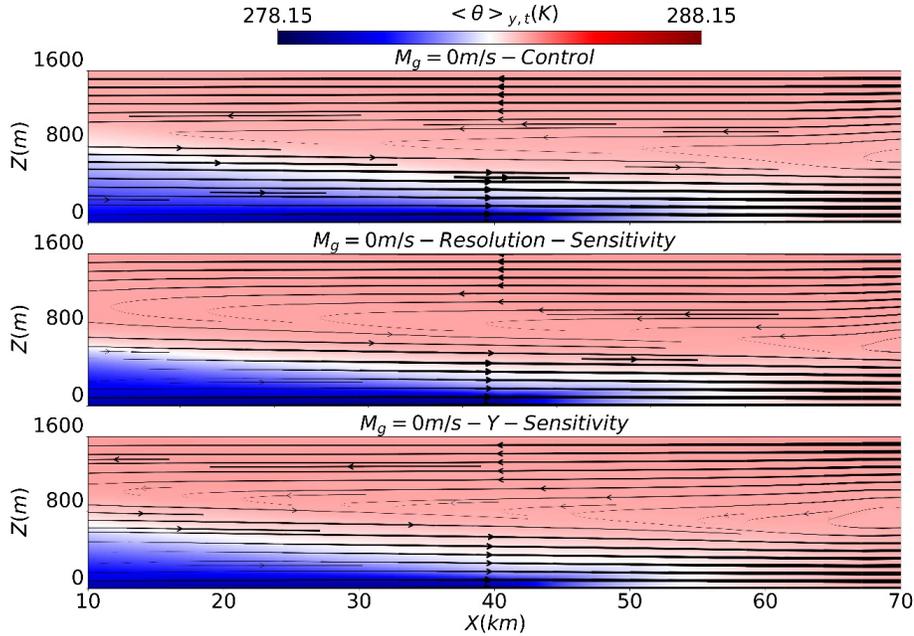

Figure S2 Pseudocolour plots of along-shore, time averaged temperature $\langle\theta\rangle_{y,t}$ through an *X-Z* slice of the analysis domain for the case of $M_g$=0 m/s (control, top), (resolution sensitivity, middle), (*Y* sensitivity, bottom) . The streamlines are drawn with varied linewidth to reflect the air speed in each subplot

Figures S-3 and S-4 show, respectively, pseudocolour plots of along-shore, time averaged streamwise velocity $\langle U \rangle_{y,t}$ and temperature $\langle\theta\rangle_{y,t}$ through an *X-Z* slice of the analysis domain (control, top), (resolution sensitivity, middle), (*Y* sensitivity, bottom) for the ($\alpha$=180° and $M_g$ =1.2 m/s) scenario. The large-scale features of the LSBs of the sensitivity simulations are here also quite similar to the control case, with more moderate differences in temperature gradients, and thus stability, near the surface.

In tables S-1 and S-2, we compare the main analysed LES outputs, but this time for the resolution and *Y* sensitivity simulations relative to the control case for the two considered scenarios: (i) $\alpha$=180° and $M_g$=1.2 m/s and (ii) $M_g$=0 m/s. The shore fluxes reported errors in table S-1 are consistently small and would not affect the modelling part and the scaling analysis.

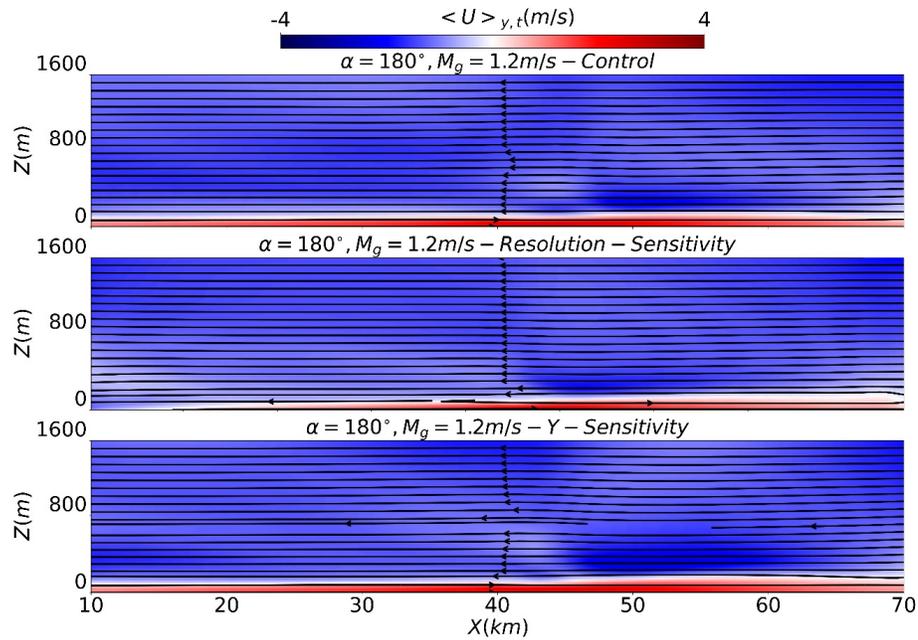

Figure S3 Pseudocolour plots of along-shore, time averaged stream-wise velocity $\langle U \rangle_{y,t}$ through an *X-Z* slice of the analysis domain for the case of $\alpha$=180° and $M_g$=1.2 m/s (control, top), (resolution sensitivity, middle), (*Y* sensitivity, bottom)

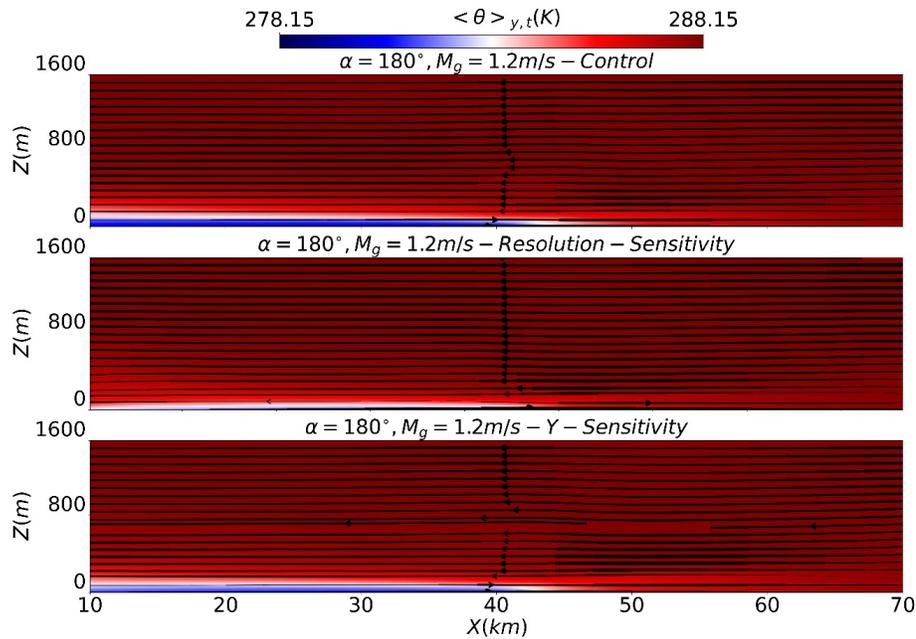

Figure S4 Pseudocolour plots of along-shore, time averaged temperature $\langle \theta \rangle_{y,t}$ through an *X-Z* slice of the analysis domain for the case of $\alpha$=180° and $M_g$=1.2 m/s (control, top), (resolution sensitivity, middle), (*Y* sensitivity, bottom). The streamlines are drawn with varied linewidth to reflect the air speed in each subplot

Table S-1 Comparing the analysed LES quantities of the resolution and Y sensitivity simulations, relative to the control simulation ($\alpha=180°$ and $M_g$ =1.2 m/s setup)

| LES outputs to compare | Simulation type | | | Simulation differences |
|---|---|---|---|---|
| | (I) Control case $M_g = 1.2$ $\alpha = 180°$ | (II) Resolution sensitivity $M_g = 1.2$ $\alpha = 180°$ | (III) Y-sensitivity $M_g = 1.2$ $\alpha = 180°$ | (IV) % $Error_{I-II}$ % $Error_{I-III}$ |
| (a) $Q_{shore}$- LES (m³/s) | −1751 | −1752 | −1736 | $Error_{I-II} \cong 0.06\%$ $Error_{I-III} \cong -0.86\%$ |
| (b) $Q_{shore}$- modeled (m³/s) | −1704 | −1704 | −1704 | $Error_{I,a-I,b} \cong -2.68\%$ $Error_{II,a-II,b} \cong -2.74\%$ $Error_{III,a-III,b} \cong -1.84\%$ |
| (a) $q_{shore}$- LES (Km/s) | −310.8 | −312.6 | −308.24 | $Error_{I-II} \cong 0.58\%$ $Error_{I-III} \cong -0.82\%$ |
| (b) $q_{shore}$-modeled (Km/s) | −301.6 | −301.6 | −301.6 | $Error_{I,a-I,b} \cong -2.96\%$ $Error_{II,a-II,b} \cong -3.52\%$ $Error_{III,a-III,b} \cong -2.15\%$ |
| $\langle w'\theta' \rangle_{Sea}$ (Km/s) | −0.0089 | −0.0107 | −0.0117 | $Error_{I-II} \cong 20.22\%$ $Error_{I-III} \cong 31.46\%$ |
| $\langle w'\theta' \rangle_{Land}$ (Km/s) | 0.0235 | 0.0143 | 0.0228 | $Error_{I-II} \cong -39.15\%$ $Error_{I-III} \cong -2.98\%$ |
| $SB_{depth}$ (m) | 101.6 | 84.2 | 101.6 | $Error_{I-II} \cong -17.13\%$ $Error_{I-III} \cong 0\%$ |
| $\bar{\epsilon}/\bar{\epsilon}_0$ | 1.8 | 1.25 | 1.99 | $Error_{I-II} \cong -30.56\%$ $Error_{I-III} \cong 10.56\%$ |

Table S-2 Comparing the analysed LES quantities of the resolution and Y sensitivity simulations, relative to the control simulation ($M_g$ =0 m/s setup)

| LES outputs to compare | Simulation type | | | Simulation differences |
|---|---|---|---|---|
| | (I) Control case $M_g = 0$ | (II) Resolution sensitivity $M_g = 0$ | (III) Y-sensitivity $M_g = 0$ | (IV) % $Error_{I-II}$ % $Error_{I-III}$ |
| $\langle w'\theta' \rangle_{Sea}$ (Km/s) | −0.0039 | −0.0038 | −0.0031 | $Error_{I-II} \cong -2.56\%$ $Error_{I-III} \cong -20.51\%$ |
| $\langle w'\theta' \rangle_{Land}$ (Km/s) | 0.0619 | 0.0577 | 0.0536 | $Error_{I-II} \cong -6.79\%$ $Error_{I-III} \cong -13.41\%$ |
| $SB_{depth}$ (m) | 812.7 | 842.1 | 812.7 | $Error_{I-II} \cong 3.62\%$ $Error_{I-III} \cong 0\%$ |
| $\bar{\epsilon}_0$ (1/s²) | $2.92 \times 10^{-5}$ | $2.73 \times 10^{-5}$ | $2.8 \times 10^{-5}$ | $Error_{I-II} \cong -6.51\%$ $Error_{I-III} \cong -4.11\%$ |

The surface fluxes are sensitive to the grid resolution, and therefore this motivates an additional analysis on the effect of time averaging on the analysed statistics, which is investigated in tables S-3 to S-5. Assume the last $\tau_f$ period in every simulation is split into two halves i.e., $\tau_f = T = T_1 + T_2$ with $T_1$ (first half) and $T_2$ (second half). Table S-3 shows that the averaging time window is generally not critical for the $Y$ sensitivity simulations. The errors show the highest sensitivity for the sea heat fluxes in the $Y$ sensitivity simulations relative to the control case as shown in table S-4. Similarly for resolution sensitivity simulations in table S-5. Most importantly, shore fluxes across all sensitivity simulations are almost not affected by the averaging time, and therefore this will not affect the modelling part and the scaling analysis.

Table S-3 Comparing the analysed LES quantities of the $Y$ sensitivity simulations over different periods ($\alpha=180°$ and $M_g=1.2$ m/s setup)

| % Error in analysed LES outputs | Simulation type and differences | | |
|---|---|---|---|
| | (I) $Y$-sensitivity simulation, over $T_1$ $M_g = 1.2$ $\alpha = 180°$ | (II) $Y$-sensitivity simulation, over $T_2$ $M_g = 1.2$ $\alpha = 180°$ | (III) $Y$-sensitivity simulation, over $T$ $M_g = 1.2$ $\alpha = 180°$ |
| $Q_{shore}$ (m³/s) | −1738.04 $\text{Error}_{II-I} \cong 0.09\%$ | −1736.47 | −1737.34 $\text{Error}_{II-III} \cong 0.05\%$ |
| $q_{shore}$ (Km/s) | −308.34 $\text{Error}_{II-I} \cong 0.03\%$ | −308.25 | −308.28 $\text{Error}_{II-III} \cong 0.01\%$ |
| $\langle w'\theta' \rangle_{Sea}$ (Km/s) | −0.0105 $\text{Error}_{II-I} \cong -10.26\%$ | −0.0117 | −0.0111 $\text{Error}_{II-III} \cong -5.13\%$ |
| $\langle w'\theta' \rangle_{Land}$ (Km/s) | 0.0226 $\text{Error}_{II-I} \cong -0.88\%$ | 0.0228 | 0.0227 $\text{Error}_{II-III} \cong -0.44\%$ |
| $SB_{depth}$ (m) | 101.6 $\text{Error}_{II-I} = 0\%$ | 101.6 | 101.6 $\text{Error}_{II-III} = 0\%$ |

Table S-4 Comparing the analysed LES quantities of the *Y* sensitivity simulations relative to the control simulation ($\alpha$=180° and $M_g$=1.2 m/s setup) over different periods

| % Error in analysed LES outputs | Simulation type and differences | | |
|---|---|---|---|
| | (I) *Y*-sensitivity over $T_1$ Control case over $T_2$ $M_g = 1.2$ $\alpha = 180°$ | (II) *Y*-sensitivity over $T_2$ Control case over $T_2$ $M_g = 1.2$ $\alpha = 180°$ | (III) *Y*-sensitivity over $T$ Control case over $T_2$ $M_g = 1.2$ $\alpha = 180°$ |
| $Q_{shore}$ | −0.74% | −0.83% | −0.78% |
| $q_{shore}$ | −0.79% | −0.82% | −0.81% |
| $\langle w'\theta'\rangle_{Sea}$ | 17.82% | 31.67% | 24.75% |
| $\langle w'\theta'\rangle_{Land}$ | −3.86% | −3.16% | −3.51% |
| $SB_{depth}$ | 0% | 0% | 0% |

Table S-5 Comparing the analysed LES quantities of the resolution sensitivity simulations relative to the control simulation ($\alpha$=180° and $M_g$=1.2 m/s setup) over different periods

| % Error in analysed LES outputs | Simulation type and differences | | |
|---|---|---|---|
| | (I) Res-sensitivity over $T_1$ Control case over $T_2$ $M_g = 1.2$ $\alpha = 180°$ | (II) Res-sensitivity over $T_2$ Control case over $T_2$ $M_g = 1.2$ $\alpha = 180°$ | (III) Res-sensitivity over $T$ Control case over $T_2$ $M_g = 1.2$ $\alpha = 180°$ |
| $Q_{shore}$ | 0.13% | 0.08% | 0.1% |
| $q_{shore}$ | 0.59% | 0.56% | 0.58% |
| $\langle w'\theta'\rangle_{Sea}$ | 11.75% | 20.01% | 15.88% |
| $\langle w'\theta'\rangle_{Land}$ | −27.1% | −39.05% | −33.07% |
| $SB_{depth}$ | −0.52% | −17.11% | −17.11% |

## 9 Thermal circulations for along-shore and oblique angles

The plots below depict the streamwise and temperature pseudocolour plots for the angles not shown in the main text. The streamlines are drawn with varied linewidth to reflect the air speed in each temperature subplot. For inflows, Y-Z neutral slices are fed into the domain at X=0 for α=270° (since Ekman rotation result in a net flow from sea to land) and at X=80 km for α = 90° (Ekman here results in a land to sea flow).

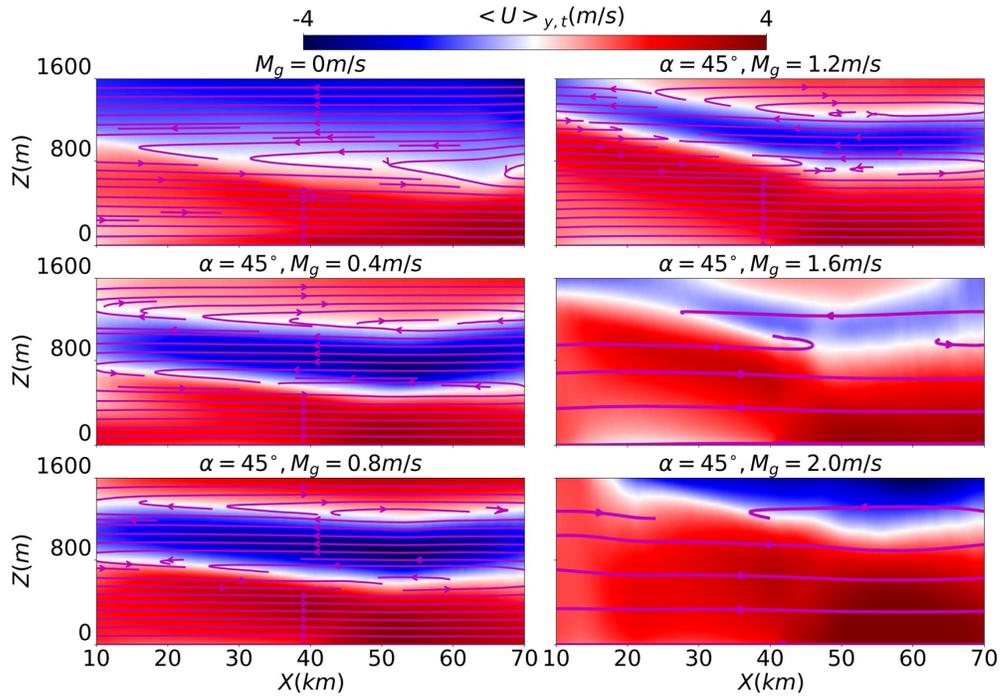

Figure S5 Pseudocolour plots of along-shore, time averaged stream-wise velocity $\langle U \rangle_{y,t}$ through an X-Z slice of the analysis domain. The series of subplots correspond to α=45° with increasing $M_g$

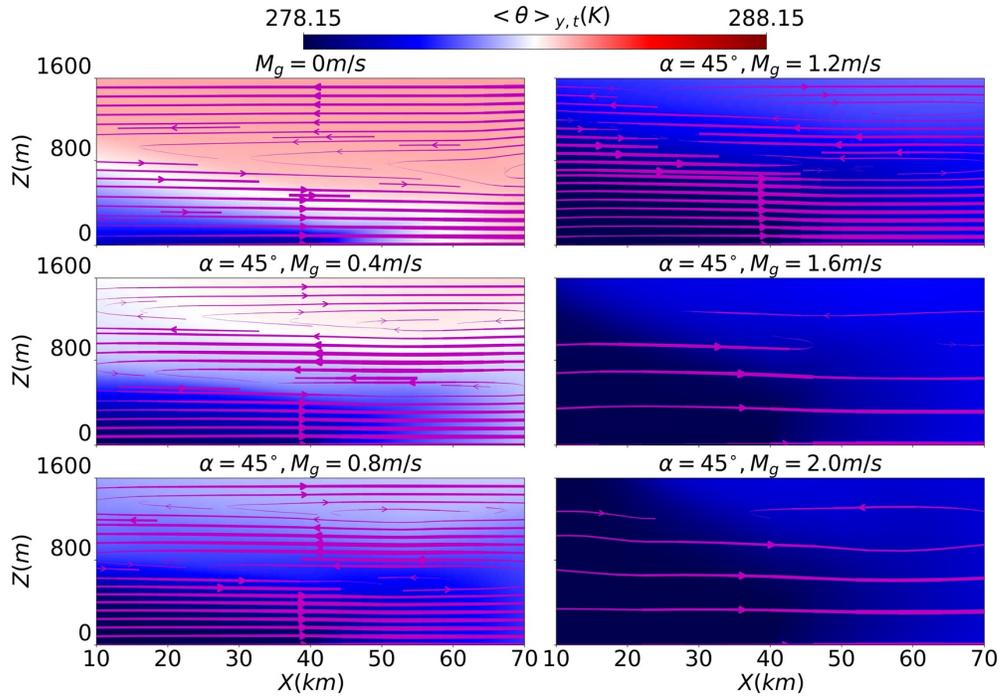

Figure S6 Pseudocolour plots of along-shore, time averaged temperature $\langle\theta\rangle_{y,t}$ through an *X-Z* slice of the analysis domain. The series of subplots correspond to $\alpha=45°$ with increasing $M_g$

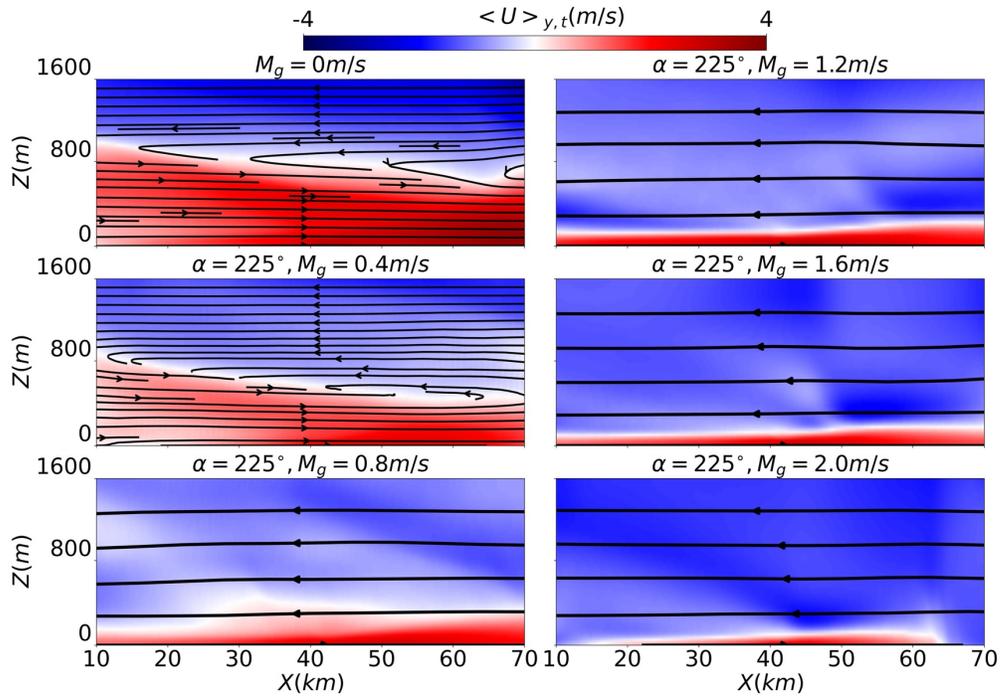

Figure S7 Pseudocolour plots of along-shore, time averaged stream-wise velocity $\langle U\rangle_{y,t}$ through an *X-Z* slice of the analysis domain. The series of subplots correspond to $\alpha=225°$ with increasing $M_g$

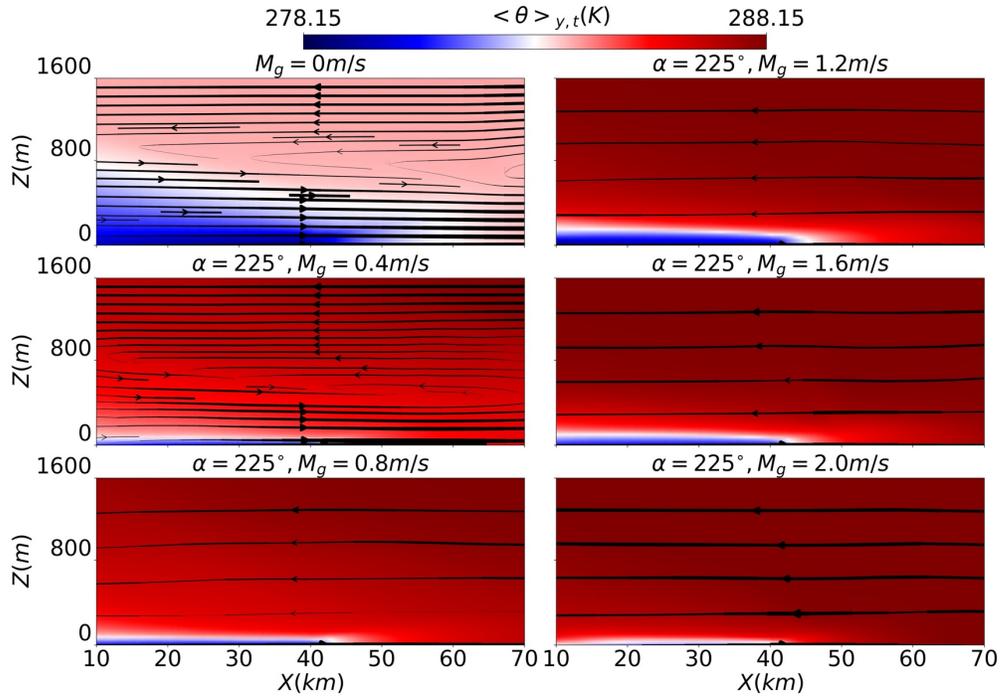

Figure S8 Pseudocolour plots of along-shore, time averaged temperature $\langle\theta\rangle_{y,t}$ through an *X-Z* slice of the analysis domain. The series of subplots correspond to $\alpha=225°$ with increasing $M_g$

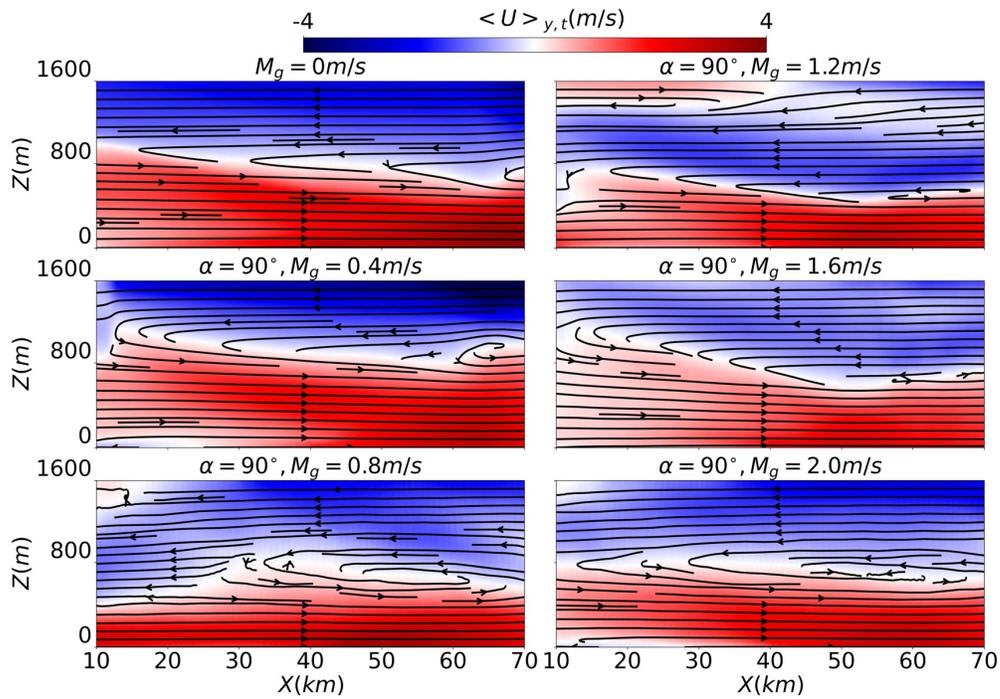

Figure S9 Pseudocolour plots of along-shore, time averaged stream-wise velocity $\langle U\rangle_{y,t}$ through an *X-Z* slice of the analysis domain. The series of subplots correspond to $\alpha=90°$ with increasing $M_g$

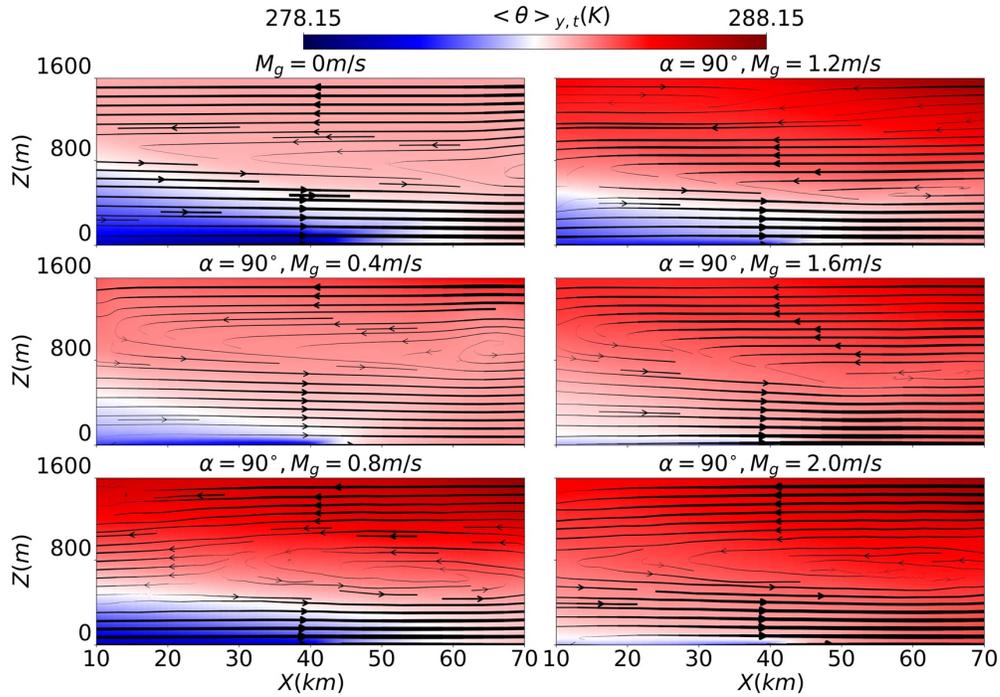

Figure S10 Pseudocolour plots of along-shore, time averaged temperature $\langle\theta\rangle_{y,t}$ through an *X-Z* slice of the analysis domain. The series of subplots correspond to $\alpha=90°$ with increasing $M_g$

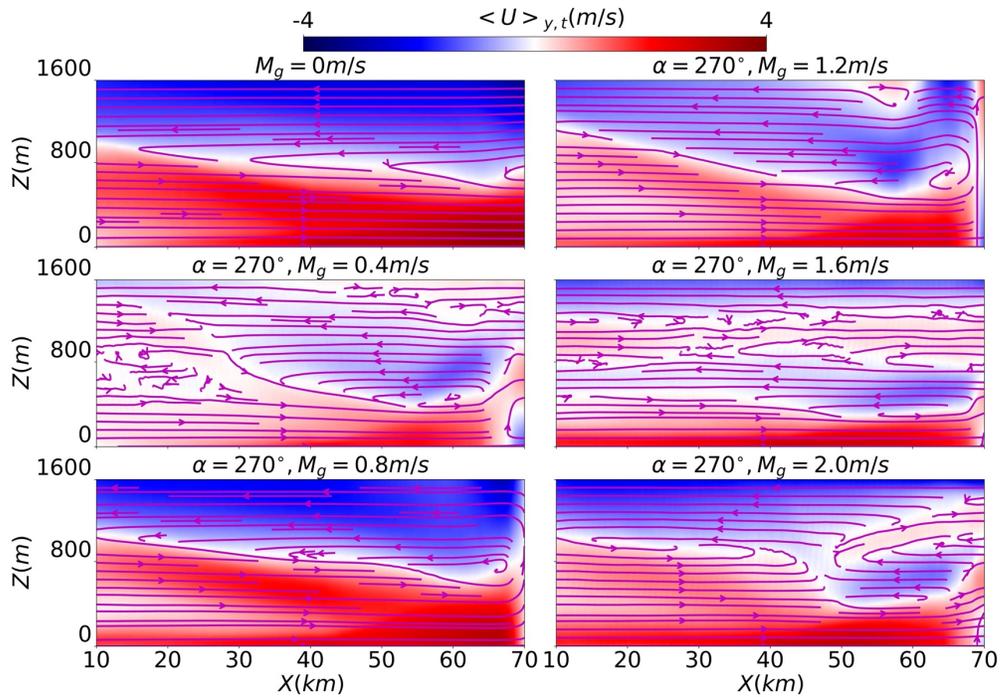

Figure S11 Pseudocolour plots of along-shore, time averaged stream-wise velocity $\langle U\rangle_{y,t}$ through an *X-Z* slice of the analysis domain. The series of subplots correspond to $\alpha=270°$ with increasing $M_g$

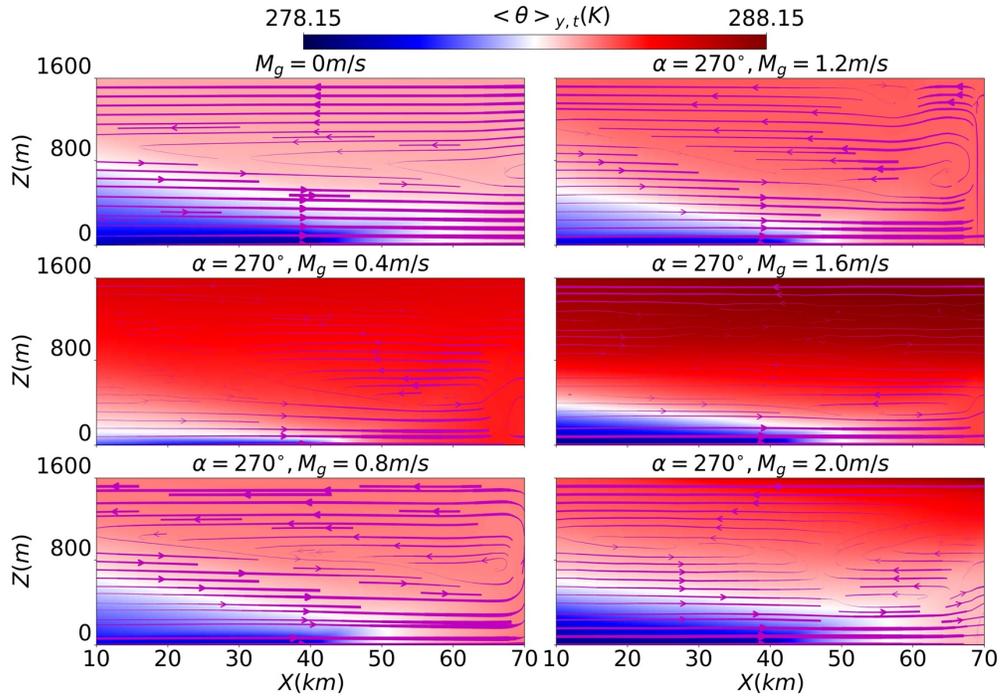

Figure S12 Pseudocolour plots of along-shore, time averaged temperature $\langle\theta\rangle_{y,t}$ through an *X-Z* slice of the analysis domain. The series of subplots correspond to $\alpha=270°$ with increasing $M_g$

## 10 Friction velocity

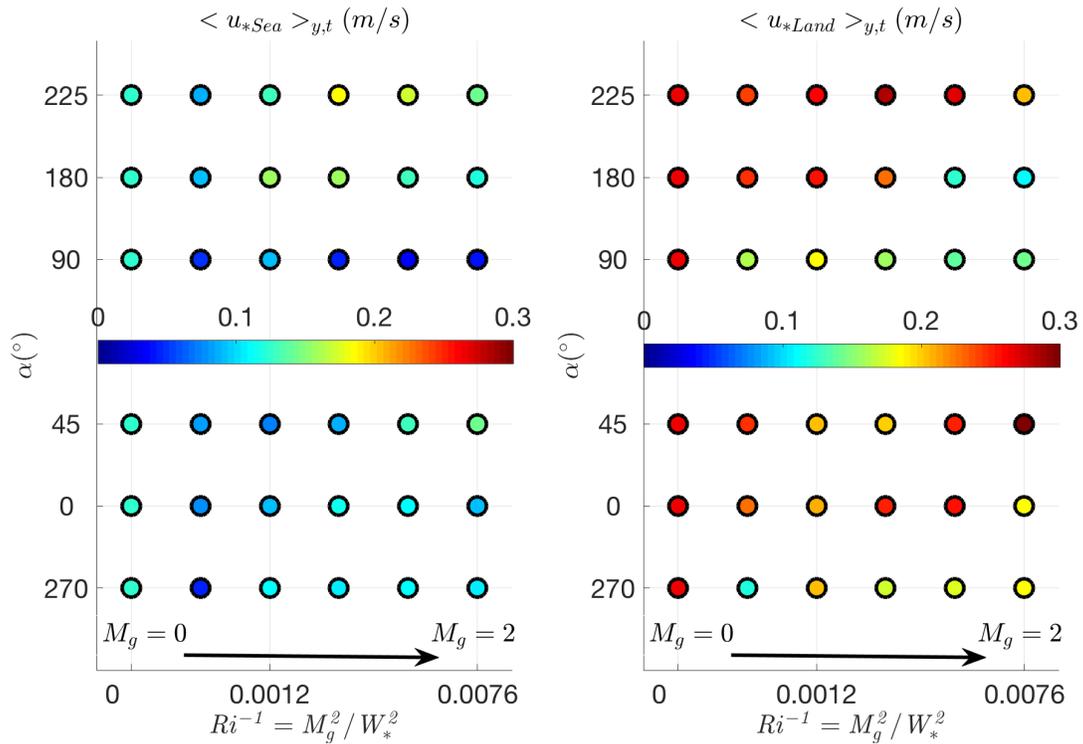

Figure S13 A scatter plot of the surface friction velocities over the two patches for all $\alpha$'s with increasing $Ri^{-1}$

(sea: left, land: right)

The rows corresponding to $\alpha$=0°, 45° and 270° in figure S-13 show that, as $M_g$ increases, $\langle u_{*Sea}\rangle_{y,t}$ decreases slightly as sea inflow neutral planes are more strongly advected. Note that the roughness lengths being equal, this difference is solely due to stability effects. The rows corresponding to $\alpha$=180° and 225° reveal that the sea friction velocity increases slightly with intermediate $M_g$ before returning to a value near the reference state of no pressure gradient state, and the initial sea friction velocity remains low along $\alpha$=90°. In the right panel of figure S-13, the row corresponding to $\alpha$=225° shows that the friction velocity over land is more resilient to perturbations of $M_g$, preserving values close to the canonical LSB reference. This is because advection of the aligned neutral land inflow planes never eliminates the near surface sea breeze dynamics. Likewise, the row corresponding to $\alpha$=180° exhibits a similar behaviour to $\alpha$=225° up to $M_g$=1.2 m/s. However, for $M_g$>1.2 m/s, $\langle u_{*Land}\rangle_{y,t}$ decreases significantly as the LSB weakens. The $\alpha$=0° and $\alpha$=45° cases exhibit much more resilience than the other rows at different $\alpha$'s. The initial $\langle u_{*Land}\rangle_{y,t}$ for $\alpha$=90° decreases more with increased $M_g$ compared to $\alpha$=270° for the same above reasons.

## 11 Streamwise advective heat flux profile

In figure S14, the reference $M_g$=0 (cyan colour) case shows two positive local peaks when $\langle U \rangle_{y,t}$>0 (within the sea breeze), and then it decreases steadily at higher levels where it shifts sign once at Z~813 m (1st layer of the return air aloft). For $M_g$=0.4 m/s and 0.8 m/s along $\alpha$=0°, the profiles $\langle U\rangle_{y,t} \langle \theta \rangle_{y,t}(z)$ change sign twice in accordance with $\langle U \rangle_{y,t}$. The profiles of $M_g$=1.2 m/s, 1.6 m/s and mainly 2 m/s are more homogenous, with mostly positive advection of heat into land (pressure driven) except near the top of the atmospheric boundary layer for $M_g$=1.2 m/s and 1.6 m/s. In the ($\alpha$=45°) subplot, the trends follow similar patterns as the ($\alpha$=0°) cases but the homogenization is delayed as $U_g$ is weakened. For both $\alpha$=0° and $\alpha$=45° the near surface advection of air from sea to land is strengthened at first and then remains significant as $M_g$ increases.

Similarly, in the ($\alpha$=180°) subplot of this figure, the profiles of all $M_g$>0 look similar. The only difference here is that as $M_g$ increases, the trends are steeper near the surface as the sea breeze depth become shallower. Likewise, the ($\alpha$=225°) subplot looks similar but less steep compared to $\alpha$=180° as $U_g$ is weakened. For both $\alpha$=180° and $\alpha$=225° the advection is weakened rapidly, monotonically, and consistently. In the ($\alpha$=270° and 90°) subplots, all the trends follow similar patterns as the initial profile.

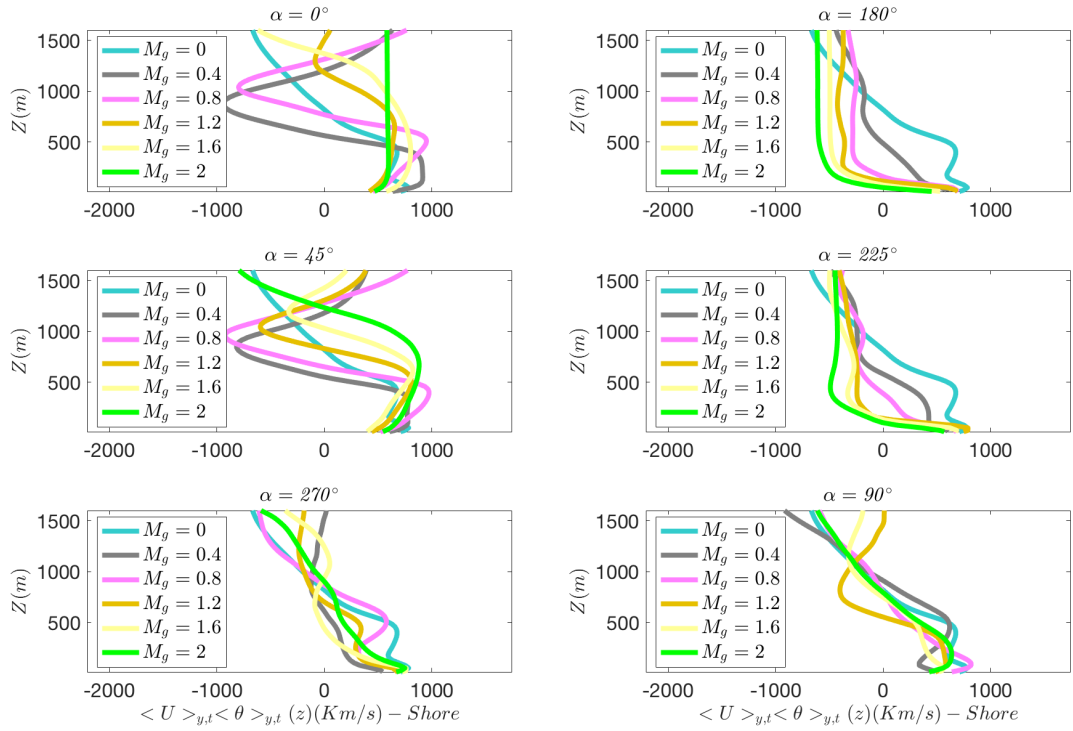

Figure S14 Shore stream-wise advective heat flux profiles for all $\alpha$'s with increasing $M_g$